\documentclass[lettersize,journal]{IEEEtran}
\IEEEoverridecommandlockouts

\usepackage{amsmath,amsfonts}
\usepackage{algorithmic}
\usepackage{algorithm}
\usepackage{array}
\usepackage[caption=false,font=footnotesize,labelfont=default,textfont=default]{subfig}
\usepackage{textcomp}
\usepackage{stfloats}
\usepackage{url}
\usepackage{verbatim}
\usepackage{graphicx}
\usepackage{cite}

\usepackage{multirow}
\usepackage{multicol}
\usepackage{xcolor}
\usepackage{booktabs}
\usepackage[nolist,nohyperlinks]{acronym}
\usepackage[scaled]{helvet}
 \begin{acronym}[XXXXXXXX]
    \acro{3gpp}[3GPP]{3rd Generation Partnership Project}
    \acro{5g}[5G]{fifth generation}
    \acro{ae-elm}[AE-ELM]{auto-encoder extreme learning machine}
    \acro{ai}[AI]{artificial intelligence}
    \acro{aoa}[AoA]{angle-of-arrival}
    \acro{aod}[AoD]{angle-of-departure}
    \acro{ap}[AP]{access point}
    \acro{awgn}[AWGN]{additive white Gaussian noise}
    \acro{ble}[BLE]{Bluetooth Low Energy}
    \acro{bw}[BW]{bandwidth}
    \acro{bwl}[BW-level]{bandwidth-level}
    \acro{bs}[BS]{base station}
    \acro{cm}[CM]{common model}
    \acro{cnn}[CNN]{convolutional neural network}
    \acro{cr}[CR]{compression ratio}
    \acro{csi}[CSI]{channel state information}
    \acro{cu}[CU]{Centralized Unit}
    \acro{dl}[DL]{downlink}
    \acro{dlaod}[DL-AoD]{downlink angle-of-departure} 
    \acro{doa}[DoA]{direction-of-arrival}
    \acro{dnn}[DNN]{dense neural network}
    \acro{du}[DU]{Decentralized Unit}
    \acro{ecdf}[ECDF]{empirical cumulative distribution function}
    \acro{eeg}[EEG]{electroencephalogram}
    \acro{ekf}[EKF]{extended Kalman filter}
    \acro{elm}[ELM]{extreme learning machine}
    \acro{em}[EM]{Encoder Model}
    \acro{fft}[FFT]{Fast Fourier Transform}
    \acro{fr}[FR]{frequency response}
    \acro{fm}[FM]{full model}
    \acro{fp}[FP]{fingerprinting}
    \acro{fr-c}[FR-Complex]{complex frequency response}
    \acro{fr-pow}[FR-Power]{power-domain frequency response}
    \acro{fr-ph}[FR-Phase]{relative phase frequency response}
    \acro{fr-pp}[FR-Power/Phase]{power and relative phase frequency response}
    \acro{fr1}[FR1]{Frequency Range 1}
    \acro{fr2}[FR2]{Frequency Range 2}
    \acro{hd}[HDim]{HIDDEN\_DIM}
    \acro{gda}[GDA]{Generalized Discriminant Analysis}
    \acro{gelu}[GELU]{Gaussian error linear unit}
    \acro{gnb}[gNB]{gNodeB}
    \acro{gnss}[GNSS]{Global Navigation Satellite System}
    \acro{ica}[ICA]{Independent Component Analysis}
    \acro{iot}[IoT]{Internet-of-Things}
    \acro{ips}[IPS]{indoor positioning system}
    \acro{kf}[KF]{Kalman filter}
    \acro{knn}[$k$-NN]{$k$-Nearest Neighbors}
    \acro{lbs}[LBS]{location-based services}
    \acro{lda}[LDA]{Linear Discriminant Analysis}
    \acro{lmf}[LMF]{Location Management Function}
    \acro{los}[LoS]{line-of-sight}
    \acro{lr}[LR]{learning rate }
    \acro{lstm}[LSTM]{long-short term memory}
    \acro{mac}[MAC]{Media Access Control}
    \acro{mdt}[MDT]{Minimization of Drive Testing}
    \acro{mee}[MEE]{mean Euclidean error}
    \acro{mimo}[MIMO]{multiple-input multiple-output}
    \acro{ml}[ML]{machine learning}
    \acro{mmw}[mmWave]{millimeter-wave}
    \acro{mrtt}[MRTT]{multi-round trip time}
    \acro{nlos}[NLoS]{non-line-of-sight}
    \acro{nn}[NN]{neural network}
    \acro{nr}[NR]{New Radio}
    \acro{ofdm}[OFDM]{orthogonal frequency-division multiplexing}
    \acro{p-aoa}[path-AoA]{path-wise angle-of-arrival}
    \acro{p-rp}[path-RP]{path-wise received power}
    \acro{prs}[PRS]{positioning reference signal}
    \acro{p-tof}[path-ToF]{path-wise time-of-flight}
    \acro{phy}[PHY]{physical-layer}
    \acro{pca}[PCA]{principal component analysis}
    \acro{pf}[PF]{particle filter}
    \acro{pl}[PL]{path-loss}
    \acro{qoe}[QoE]{quality of experience}
    \acro{qos}[QoS]{quality of service}
    \acro{rb}[RB]{resource block}
    \acro{re}[RE]{resource element}
    \acro{relu}[ReLU]{rectified linear unit}
    \acro{rbl}[RB-level]{resource block-level}
    \acro{rf}[RF]{radio frequency}
    \acro{rfid}[RFID]{radio requency identification device}
    \acro{rms}[RMS]{root mean square}
    \acro{rmsee}[RMSEE]{root mean squared euclidean error}
    \acro{rnn}[RNN]{recurrent neural network}
    \acro{rp}[RP]{reference point}
    \acro{rrc}[RRC]{radio resource control}
    \acro{rs}[RS]{recommender system}
    \acro{rsrp}[RSRP]{reference signal received power}
    \acro{rsrq}[RSRQ]{reference signal received quality}
    \acro{rss}[RSS]{received signal strength}
    \acro{rssi}[RSSI]{received signal strength indicator}
    \acro{rtt}[RTT]{round trip time}
    \acro{rx}[RX]{receiver}
    \acro{slam}[SLAM]{simultaneous localization and mapping}
    \acro{sp}[SP]{scattering point}
    \acro{ss}[SS]{synchronization signal}
    \acro{ssb}[SSB]{synchronization signal block}
    \acro{srs}[SRS]{sounding reference signal}
    \acro{sss}[SSS]{secondary synchronization signal}
    \acro{std}[std]{standard deviation}
    \acro{svm}[SMV]{support vector machine}
    \acro{svd}[SVD]{single value decomposition}
    \acro{tdoa}[TDoA]{time-difference-of-arrival}
    \acro{tl}[TL]{transfer learning}
    \acro{tlcnn}[TLCNN]{Transfer Learning Convolutional Neural Network}
    \acro{toa}[ToA]{time-of-arrival}
    \acro{tof}[ToF]{time-of-flight}
    \acro{tsne}[t-SNE]{T-distributed Stochastic Neighbor Embedding}
    \acro{tx}[TX]{transmitter}
    \acro{ue}[UE]{user equipment}
    \acro{ul}[UL]{uplink}
    \acro{ulaoa}[UL-AoA]{uplink angle-of-arrival}
    \acro{wifi}[\mbox{Wi-Fi}]{IEEE 802.11 network}
    \acro{wlan}[WLAN]{wireless LAN}
    \acro{wsn}[WSN]{wireless sensors networks}
  \end{acronym}
\usepackage{color}
\usepackage{soul}
\usepackage{siunitx}
\usepackage{bm}
\usepackage{array}
\usepackage{orcidlink}
\usepackage[export]{adjustbox}
\usepackage{microtype}
\usepackage{balance}
\usepackage{multirow}
\usepackage{hyperref}

\newcommand{\blue}[1]{{\color{black}#1}}
\newcommand{\red}[1]{{\color{black}#1}}

\renewcommand{\vec}[1]{\bm{#1}}

\def\BibTeX{{\rm B\kern-.05em{\sc i\kern-.025em b}\kern-.08em
    T\kern-.1667em\lower.7ex\hbox{E}\kern-.125emX}}

\usepackage[font = footnotesize, textfont=up]{caption}
    \usepackage{etoolbox}
\makeatletter 
\pretocmd\@bibitem{\color{black}\csname keycolor#1\endcsname}{}{\fail}
\newcommand\citecolor[1]{\@namedef{keycolor#1}{\color{blue}}}
\makeatother

\begin{document}

\title{Robust NLoS Localization in 5G mmWave Networks: Data-based Methods and Performance}

\author{
    \IEEEauthorblockN{Roman Klus\,\orcidlink{0000-0002-0641-5931},~\IEEEmembership{Member,~IEEE,}
    Jukka Talvitie\,\orcidlink{0000-0001-7685-7666}, 
    Julia Vinogradova\,\orcidlink{0000-0001-8911-2065},
    Gabor Fodor\,\orcidlink{0000-0002-2289-3159},
    Johan Torsner,
    and \\Mikko Valkama\,\orcidlink{0000-0003-0361-0800},~\IEEEmembership{Fellow,~IEEE}} 
    \vspace{-1mm}

    \thanks{Limited subset of early-stage results presented at {IEEE SPAWC 2022~{[14]}}.}
    \thanks{R. Klus, J. Talvitie, and M. Valkama are with Tampere University, Finland. 
    }
    \thanks{J. Vinogradova and J. Torsner are with Ericsson Research, Helsinki, Finland. 
    }
    \thanks{G. Fodor is with Ericsson Research and with KTH, Stockholm, Sweden. 
    }
    \thanks{\blue{Data and codes openly available at \url{https://doi.org/10.5281/zenodo.12204893}}}
    }

\maketitle
\begin{abstract}
Ensuring smooth mobility management while employing directional beamformed transmissions in 5G millimeter-wave networks calls for robust and accurate \ac{ue} localization and tracking.  
In this article, we develop neural network-based positioning models with time- and frequency-domain {\ac{csi}} data in harsh \ac{nlos} conditions. We propose a novel frequency-domain feature extraction, {which combines} relative phase differences and received powers across resource blocks, {and offers} robust performance and reliability. Additionally, we exploit the multipath components and propose an aggregate time-domain feature combining time-of-flight, angle-of-arrival and received path-wise powers. Importantly, {the temporal correlations are also} harnessed in the form of sequence processing neural networks, {which prove} to be of particular benefit for vehicular UEs. Realistic numerical evaluations in large-scale \ac{los}-obstructed urban environment with {moving vehicles} are provided, building on full ray-tracing based propagation modeling. The {results} show the robustness of the proposed \ac{csi} features in terms of positioning accuracy, {and that} the proposed models reliably {localize} \acp{ue} {even in the absence of} a \ac{los} path, clearly outperforming the state-of-the-art \blue{with similar or even reduced processing complexity}. The proposed sequence-based {neural network} model is capable of {tracking} the \ac{ue} position, speed and heading simultaneously despite the strong uncertainties in the \ac{csi} measurements. \red{Finally, it is shown that differences between the training and online inference environments can be efficiently addressed and alleviated through transfer learning.}
\end{abstract}
\vspace{-1mm}

\acresetall 
\vspace{-0.5mm}
\vspace{-5mm}
\section{Introduction}
\label{sec:1}
\acresetall
\IEEEPARstart{E}{xpanding} to the \ac{mmw} frequencies allows {to harness} large channel bandwidths in the \ac{5g} \ac{nr} mobile communication networks, which improves the network capacity, peak data rates, and latency characteristics compared to legacy systems \cite{ToskalaBook, 2020Dahlman5G}. In such \ac{mmw} networks, beamforming active antenna arrays are a critical technology, allowing directional transmission and reception capabilities, thus improving the link budget while mitigating co-channel interference and providing the basis for angle-based \emph{cellular positioning}.

In general, accurate real-time knowledge of the locations of the network \ac{ue} is critical to ensure smooth and seamless mobility management, efficient handover management, and improved reliability of the radio link, while also allowing for \ac{lbs} \cite{SituationalAware_2019, del2017survey,7984759, LocaAware_Maga_Henk_2014}. Baseline \ac{ue} localization builds commonly on \ac{gnss} based approaches. {However,} terrestrial positioning utilizing \ac{5g} and other signals of opportunity is of increasing importance, and is also the main technical scope of this article. This is well motivated, as the availability of \ac{gnss} is known to be compromised not only indoors but also in outdoor urban areas~\cite{del2017survey,nagai2020evaluating}, while the large channel bandwidths and directional antenna systems deployed in \ac{5g} allow for accurate time- and angle-based measurements.

There is generally a wide selection of positioning methods 
available in the literature~\cite{del2017survey}, covering both model-based Bayesian filtering approaches {\cite{10086654, ge2022computationally, villacres2019particle, koivisto2017joint, ko2022high, talvitie2017novel}} as well as data-driven \ac{ml} based methods {\cite{klus2022machine, 8823059, zappone2019wireless, revach2022kalmannet, klus2021neural, torres2020comprehensive, rojo2019machine, klus2021transfer, butt2021ml, gonultacs2021csi, wang2016csi, chen2017confi, xiao2012fifs, wang2016csiphase, ferrand2020dnn, gao2022towards, lynch2020localisation, hajiakhondi2021bluetooth}}.
Majority of the works focus on the \ac{los} scenarios, where the channel includes a direct propagation path from the \ac{tx} to the \ac{rx}, and thus the \ac{ue} location can be estimated geometrically by utilizing pseudo-ranges and/or angular information based on radio measurements. Good examples of such \ac{los}-oriented 5G positioning works include {\cite{koivisto2017joint, ko2022high}}, building commonly on Bayesian filtering methods such as different variants of \ac{ekf} and \ac{pf}. In case the \ac{los} path is unavailable, geometry-based Bayesian filtering models still exist, e.g.,~{\cite{ge2022computationally, villacres2019particle, talvitie2017novel}}.  Such methods are, however, typically limited to single-bounce scenarios with a single path per scattering point, and can thus easily become unreliable in realistic scattering environments while being also computationally {heavy and complex~\cite{9832776,feigl2021robust}}. Therefore, \ac{ue} positioning and tracking in \emph{\ac{nlos} or \ac{los}-ambiguous scenarios} call for novel solutions and models, capable of ensuring fast and reliable operation in realistic scattering environments with feasible real-time computational complexity. This is the main technical focus of this article, with a specific emphasis on vehicular systems in challenging urban environments, \red{where \ac{nlos} scenarios commonly occur with realistic network deployments \cite{3GPPchannel}}.

In this article, building on our initial work in \cite{klus2022machine}, we propose and describe efficient \ac{ml} based models for \ac{nlos} or \ac{los}-obstructed positioning that {offer} robust performance, low operational complexity, good generalization properties, and wide architectural options. We harness the temporal correlation of the channel features in vehicular systems and focus on sequence processing \ac{nn} methods as the fundamental ML engine. \blue{We propose two alternative feature sets capable of describing the radio channel and further \ac{nn}-based positioning, namely \emph{frequency-domain and time-domain \ac{csi} features}}. Additionally, we include the vehicle speed and heading as additional model outputs to enable efficient  tracking using a single model.

\ac{ml}-based positioning has been addressed in the recent literature, e.g. in {\cite{8823059, klus2022machine, zappone2019wireless, revach2022kalmannet, klus2021neural, rojo2019machine, butt2021ml, gonultacs2021csi, wang2016csi, chen2017confi, xiao2012fifs, wang2016csiphase, ferrand2020dnn, klus2021transfer, gao2022towards, lynch2020localisation, hajiakhondi2021bluetooth}}. To this end, \ac{rss} measurements were adopted in~\cite{torres2020comprehensive, rojo2019machine, klus2021transfer}, in indoor \ac{wifi} fingerprinting context, while the corresponding \ac{5g} deployment was considered in~\cite{butt2020rf}. 
The work in \cite{wang2016csi} proposed a CSI-based fingerprinting model that utilizes the amplitude response of the channel. 
Due to the classifier-like \ac{nn}, the proposed system is, however, limited to small deployments. 
The authors of~\cite{8823059} developed an \ac{nn}-based feature extractor with \ac{wifi} CSI measurements considering the amplitude response only, followed by a \ac{knn} positioning algorithm. 
Similar approaches building on channel amplitude response measurements were considered also in~\cite{chen2017confi} and~\cite{xiao2012fifs}. In \cite{wang2016csiphase}, a \ac{wifi} positioning approach utilizing the channel phase response as a feature was proposed. 
The method is, however, not suitable for large-scale scenarios 
due to the classifier-based \ac{nn}, while the considered phase slope calculation is also subject to ambiguities. 
The work in \cite{butt2021ml}, in turn, presented a 5G positioning system, however, being
limited to the \ac{los} scenario while utilizing only the beam-specific \ac{rsrp} values as the features. In \cite{gonultacs2021csi}, a positioning system that generates probability maps using an \ac{nn} model was described with a feature representation that transforms the frequency-domain \ac{ul} \ac{csi} data to a delay-domain. 
The probability maps enable efficient sensor fusion, 
yet their scale directly affects the complexity of the underlying \ac{nn} model.
\textls[-1]{Furthermore, \cite{gao2022towards} described a paradigm to produce a high-accuracy \ac{5g} localization dataset, building on channel frequency response measurements.} 

Different hybrid solutions also exist in the literature, either in terms of aiding Bayesian filtering solutions through ML methods or fusing measurements from various different sensors~\cite{zappone2019wireless}. To this end, \cite{revach2022kalmannet} proposed a series of \ac{rnn} models to replace the \ac{ekf} and thus enable implicit and data-driven learning. 
Furthermore, a sensor fusion approach using reinforcement learning-assisted particle filtering is described in~\cite{villacres2019particle}. 
Neural network based 5G fingerprinting and \ac{gnss} data fusion were, in turn, considered in~\cite{klus2021neural}. Recently, in~\cite{chen2022joint}, an \ac{nn} model classifying different propagation paths from time-domain \ac{csi} in the form of path parameters was paired with geometry-based positioning algorithm considering \ac{los} and single-bounce \ac{nlos} paths. 

\ac{ml}-based localization with propagation time measurements has also gained interest in recent works~\cite{lynch2020localisation, feigl2021robust, klus2022machine}. 
In~\cite{lynch2020localisation}, a bidirectional \ac{rnn} is employed to track the \ac{ue} based on the \ac{toa} measurements from multiple nodes, while the authors of~\cite{feigl2021robust} utilize a \ac{cnn} to estimate the accurate \ac{toa} from the raw channel impulse responses in \ac{los}/\ac{nlos} scenarios. However, only \ac{los} measurements are used for the actual localization.
Finally, angular information at either the \ac{gnb} or the \ac{ue} can also be used for \ac{ml}-based localization, {as shown} 
in~\cite{hajiakhondi2021bluetooth}.

\begin{table*}[t]
\caption{\textsc{Summary of related works}}
\label{tab:literature}
    \centering
    \begin{tabular}{c|c|c|c|c|c|c|c|c|l}
    \toprule
    {\multirow{ 2}{*}{\textbf{Reference}}} & 
    {\multirow{ 2}{*}{\textbf{Technology}}} & 
    \multirow{ 2}{*}{\textbf{Features}} & 
    \multirow{ 2}{*}{\textbf{ML model}} & 
    \multirow{ 2}{*}{\,\textbf{Tracking}} & 
    \multicolumn{3}{c}{\textbf{\red{Uncertainty}}} & \textbf{\red{Vehicular}} & 
    \multirow{ 2}{*}{\,\textbf{NLoS}} \\ 
     &  &  &  &  & 
    {\textbf{\red{Feat.}}} & {\textbf{\red{Label}}} & {\textbf{\red{Envir.}}} & 
    {\textbf{\red{Systems}}}
    & \\ 
    \midrule

{\cite{klus2022machine}}	&	5G mmWave	&	AoA and ToF	&	DNN, LSTM	&	$\checkmark$	& $\checkmark$ & $\checkmark$ &	$\times$	&	$\checkmark$	& $\checkmark$ \vspace{-0.05mm}\\
{\cite{	8823059	}}	&	{\ac{wifi}}	&	{channel amplitude response}	&	CNN	&	$\times$	& $\times$& $\times$&	$\times$ &	$\times$	&	~$\checkmark$*	\vspace{-0.05mm}\\
\cite{	klus2021neural	}	&	5G \ac{mmw}	&	RSRP	&	DNN	&	$\times$	& $\checkmark$ &$\times$&	$\times$ &	$\times$	&	~$\times$	\vspace{-0.0mm}\\
\cite{	butt2021ml	}	&	5G \ac{mmw}	&	RSRP	&	DNN	&	$\times$	& $\times$ & $\times$ &	$\times$ &	$\times$	&	~$\times$**	\vspace{-0.0mm}\\
\cite{	gonultacs2021csi	}	&	\ac{wifi}	&	channel frequency response	&	DNN	&	$\times$	& $\checkmark$ & $\times$ &	$\times$ &	$\times$	&	~$\checkmark$	\vspace{-0.0mm}\\

\cite{	wang2016csi	}	&	\ac{wifi}	&	channel amplitude response	&	DNN	&	$\checkmark$	& $\times$ & $\times$ &	$\times$ &	$\times$	&	~$\times$	\vspace{-0.05mm}\\
\cite{	chen2017confi	}	&	\ac{wifi}	&	channel amplitude response	&	CNN	&	$\times$	& $\times$ & $\times$ &	$\times$ &	$\times$	&	~$\times$*	\vspace{-0.0mm}\\
\cite{	xiao2012fifs	}	&	\ac{wifi}	&	channel amplitude response	&	ML	&	$\checkmark$	& $\times$ & $\times$ &	$\times$ &	$\times$	&	~$\times$*	\vspace{-0.0mm}\\
\cite{	wang2016csiphase	}	&	\ac{wifi}	&	channel phase response	&	DNN	&	$\checkmark$	& $\times$ & $\times$ &	$\times$ &	$\times$	&	~$\times$	\vspace{-0.0mm}\\
\cite{	gao2022towards	}	&	5G \ac{mmw}	&	channel frequency response	&	CNN	&	$\times$	& $\checkmark$ & $\times$ &	$\times$ &	$\times$	&	~$\times$	\vspace{-0.0mm}\\
\cite{	lynch2020localisation	}	&	Unspecified	&	ToA	&	RNN	&	$\checkmark$	& $\times$ &	$\times$ &	$\times$ &	$\times$	&	~$\times$	\vspace{-0.0mm}\\
\cite{	hajiakhondi2021bluetooth	}	&	BLE	&	AoA	&	CNN	&	$\times$	& $\checkmark$ & $\times$ &	$\times$ &	$\times$	&	~$\checkmark$***	\vspace{-0.0mm}\\
\cite{	feigl2021robust	}	&	LTE	&	channel impulse response	&	CNN	&	$\times$	& $\checkmark$ & $\times$ &	$\times$ &	$\times$	&	~$\checkmark$*	\vspace{-0.0mm}\\
\cite{	chen2022joint	}	&	\ac{mmw}	&	path-wise \ac{csi} 	&	DNN	&	$\times$	& $\checkmark$ & $\times$ &	$\times$ &	$\checkmark$	&	~$\checkmark$**	\vspace{-0.0mm}\\
\,This Work\,	&	\,5G \ac{mmw}\,	&	\,time- and frequency-domain \ac{csi}\, 	&	\,DNN, LSTM\,	&	$\checkmark$	& $\checkmark$ & $\checkmark$ &	$\checkmark$ &	$\checkmark$	&	~$\checkmark$	\\
\bottomrule

\end{tabular}
\\
\vspace{1mm}
{ *\,Some NLoS samples available in evaluation;}
{ **\,Removing the detected \ac{nlos} samples before positioning;}
{ ***\,RX signal subject to Rayleigh fading.}
\vspace{-2mm}
\end{table*}

The main related works and their relevant aspects are summarized in Table~\ref{tab:literature}. Importantly, the \ac{nlos} positioning under rich and realistic scattering is not explicitly addressed, in particular in the context of beamforming \ac{5g} \ac{mmw} networks. 
Thus, complementary to the existing literature, this article focuses on \ac{ml}-based reliable network localization using time- and frequency-domain \ac{csi} data with emphasis on challenging \ac{nlos} scenarios, \red{while noting also various relevant uncertainty aspects}. The application focus is on vehicular systems in urban environments with \ac{5g} \ac{mmw} deployments and \ac{csi} features that can be obtained through 3GPP 
standardized \ac{ul} and/or \ac{dl} measurements and corresponding signaling. The contributions and novelty compared to the existing \ac{ml}-based positioning literature can be stated and summarized as follows:
\begin{itemize}
    \item We introduce, derive, and evaluate efficient frequency-domain \ac{csi} features in the form of sparse power and phase measurements, and their combinations, for \ac{ml}-based positioning models and compare their robustness with the ones {introduced previously} in the literature;
    \item \blue{We also introduce and evaluate alternative time-domain path-wise \ac{csi} features} and demonstrate their effectiveness in wireless positioning scenarios while finding the relevant and best-performing feature combinations;
    \item We develop a novel hybrid \ac{nn} processing model in terms of instantaneous and sequence data processing for simultaneous \ac{ue} location, velocity, and heading tracking using the above channel-based features;
    \item We evaluate the performance of different features and processing models in a large-scale, realistic, urban scenario with full ray-tracing-based channel measurements under harsh \ac{nlos} conditions in the context of 28\,GHz \ac{mmw} \ac{5g} network, while also \red{considering realistic training and measurement uncertainties};
    \item We show 
    that the proposed time-domain and frequency-domain features outperform the benchmark solutions, especially when combined with the sequence processing \ac{ml} model, in terms of the positioning accuracy and complexity -- in particular, in the challenging multi-bounce scattering environments, where the proposed positioning approach achieves comparable accuracy in both \ac{los} and \ac{nlos} {conditions}, regardless of the number of bounces;
    \red{\item We also address the important practical issue of environment or \ac{gnb} deployment differences between the training phase and the actual online inference phase, through transfer learning, and show that specializing to the current deployment is feasible;}
    \item \blue{Finally, we address the complexity of the developed methods, in comparison to the prior-art, while also openly share the data and codes for research reproducibility and transparency.}
\end{itemize}

\red{For clarity it is stated that selected time-domain features were initially considered in \cite{klus2022machine}, however, the adopted \ac{nn} models were lacking the advanced sequence processing capabilities. Additionally, no frequency-domain features were considered, while the transfer learning aspects were also fully neglected.} 

The rest of this article is organized as follows. 
Section~\ref{sec:3} introduces the network measurements applicable for positioning and their acquisition with 3GPP compatible reference signals and measurement procedures. Section~\ref{sec:4} introduces the proposed frequency-domain and time-domain \ac{csi} data features and the related pre-processing. Additionally, the \ac{nn} processing models and architectures are described incorporating both instantaneous and sequence models. Section~\ref{sec:5} describes the considered urban vehicular positioning scenario and evaluation environment, together with the practical measurement or data uncertainties. Additionally, the obtained numerical results are presented and analyzed, \red{while also considering the important aspect of specializing to the prevailing gNB deployment through transfer learning}. Finally, Section~\ref{sec:6} concludes the work.
\vspace{-1mm}
\section{\red{Positioning Measurements and Data Acquisition}}
\label{sec:3}
\begin{figure*}[t]
    \centering
    \includegraphics[width=1\textwidth,trim={0cm 0.3cm 0cm 0.2cm},clip] {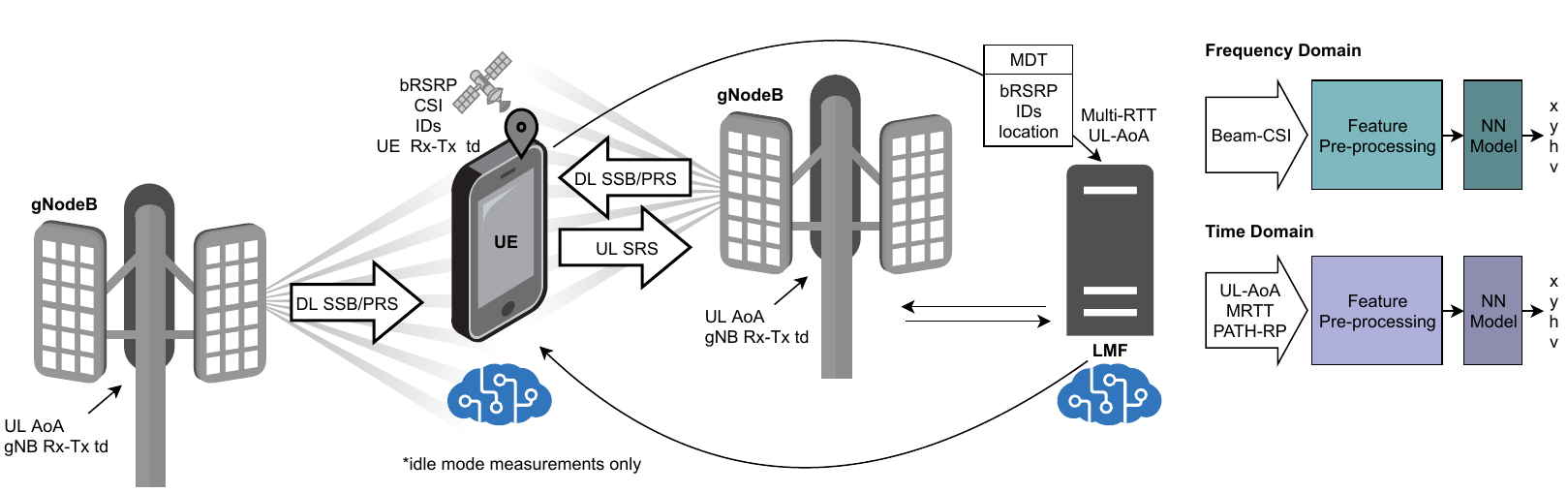} 
    \vspace{-4mm}
    \caption{{Illustration of the network data acquisition scheme with two \acp{gnb}, a \ac{ue}, and a network localization entity represented by the \ac{lmf}. Different \ac{ul} and \ac{dl} reference signals form the physical basis for obtaining positioning measurements and data.} \blue{Additionally, the baseline neural processing chains from features to UE location, heading and velocity are highlighted, for both time- and frequency-domain feature scenarios.}
    }
    \label{fig:dacq}
\end{figure*}

\vspace{-0.5mm}
\textls[-1]{This section introduces the \blue{signals and standardized measurements, available in \ac{5g} \ac{nr}, to extract positioning data}. Specific focus is on \ac{ssb} based measurements in \ac{dl}, in terms of frequency-domain data, while in time-domain we harness \ac{mrtt} and \ac{ul} \ac{aoa} based multipath measurements utilizing \ac{ul} \ac{srs} and \ac{dl} \ac{prs}. \blue{The relevant measurements and data acquisition methods are illustrated conceptually in Fig.~\ref{fig:dacq}, while being described in detail below.} \blue{For clarity, we state that the frequency- and time-domain measurements are \emph{alternative} approaches to obtain positioning features and data.}}

\vspace{-2mm}
\subsection{Signals and Measurements for 5G NR Positioning}
\label{sec:3basics}
\vspace{-0.5mm}
Measuring the received signal strength is one common approach utilized in wireless positioning. 
In \ac{5g} \ac{nr}, we distinguish \ac{rsrp}, \ac{rsrq}, and \ac{rssi}, including their beam-specific and resource-specific alternatives. Their acquisition 
is defined 
in~\cite{3GPPphy} building on different reference signals, such as \ac{ul} \ac{srs} and \ac{dl} \ac{ss} and \ac{prs}. 
The corresponding measurements are called \ac{ul}-\ac{srs}-\ac{rsrp}, \ac{ss}-\ac{rsrp} and \ac{dl}-\ac{prs}-\ac{rsrp}, respectively. Signal strength measurements are vital for numerous network functions, such as mobility management, and thus regularly collected.
 
Compared to signal strength-based measurements, propagation delay or \ac{tof}-based ranging benefits from large transmission bandwidths while being less sensitive to channel effects, such as reflections, diffractions, and scattering. 
To relax clock synchronization requirements between \ac{tx} and \ac{rx}, the \ac{5g} \ac{nr} standard supports \ac{mrtt} measurements~\cite{3GPPue} 
where the \ac{gnb} measures the round trip time, denoted as "gNB Rx--Tx time difference", based on \ac{prs} transmission in \ac{dl} and \ac{srs} transmission in \ac{ul}. In addition, the \ac{ue} measures the time between receiving the \ac{prs} and sending the \ac{srs}, denoted as the "UE Rx--Tx time difference", which is reported to the \ac{gnb}~\cite{3GPPphy} in order to solve the channel-dependent propagation delay. 
Alternatively, the \ac{tof} can be estimated indirectly at the \ac{gnb} via \ac{tdoa} and the related positioning calculations, as defined in~\cite{3GPPue}. Obtaining the \ac{tof} directly at the \ac{ue} 
is currently not explicitly standardized.

\textls[-3]{Beamformed radio access provides inherent support for angle estimation and corresponding angle-based positioning schemes. In the current \ac{5g} \ac{nr} standard, angle estimation is directly specified only for the \ac{gnb}-side angular information either via the \ac{ulaoa} or the \ac{dlaod} \cite{3GPPue}. The \ac{ulaoa} is defined as the estimated azimuth and vertical angles of a \ac{ue}, observed at a \ac{gnb}~\cite{3GPPphy}, based on \ac{ul} \ac{srs}. The exact angle estimation method used in the \ac{ulaoa} is not specified, which allows for performance optimization. 
On the other hand, the estimation of the \ac{gnb} angle at the \ac{ue} using the \ac{dlaod} is practically restricted to the use of spatial power measurements, i.e., \ac{dl}-\ac{prs}-\ac{rsrp}, which limits the achievable angle estimation accuracy~\cite{3GPPue}.}

Importantly, estimating the ranges and angles also for paths beyond the \ac{los} component is feasible~\cite{ talvitie2017novel}, 
offering added value to the positioning task~\cite{ ge2022computationally}. Since the current \ac{5g} \ac{nr} standard does not specify accurate estimation of path ranges or angles at the \ac{ue} side, for a \ac{dl}-based positioning we consider observing frequency-domain \ac{csi} measurements based on \ac{ss}s transmissions by the \ac{gnb}s. This is beneficial since the \ac{ss}s are periodically transmitted and thus systematically available. 
Additionally, for an \ac{ul}-based positioning method, we consider observing time-domain multipath measurements, including path-wise angles and propagation delays, directly at the \ac{gnb}s. For obtaining the path-wise angles and propagation delays in practice, it is possible to exploit the above-discussed \ac{5g} \ac{nr} specified \ac{ulaoa} and \ac{mrtt} methods, respectively. 

\vspace{-2mm}
\subsection{Frequency-Domain Channel Measurements through Beamformed DL SSs}
\textls[-2]{The frequency-domain \ac{csi} relates to the \ac{fr} of the effective channel between the \ac{gnb} and the \ac{ue}. Considering \ac{ofdm} based transmission, the antenna-element-wise \ac{fr} at subcarrier $n$, denoted as $\boldsymbol{\mathcal{H}}(n)\! \in \!\mathbb{C}^{N_{\text{RX}} \times N_{\text{TX}}}$, can be written as \cite{talvitie2017novel,ge2022computationally}}
\begin{equation}
\label{eq:frequency_response}
  \boldsymbol{\mathcal{H}}(n) = \sum_{k=0}^{K-1} h_k e^{-\frac{j2\pi n \tau_k F_s}{N}} \vec{a}_{\text{RX}}(\theta_{\text{AOA},k}) \vec{a}_{\text{TX}}^H(\theta_{\text{AOD},k}),
\end{equation}
\textls[-2]{where $K$ is the number of paths, while $h_k$, $\tau_k$, $\theta_{\text{AOA},k}$ and $\theta_{\text{AOD},k}$ are the complex path coefficient, \ac{tof}, \ac{aoa} and \ac{aod} for the $k^\text{th}$ path, respectively. Furthermore, $N_{\text{TX}}$ and $N_{\text{RX}}$ are the numbers of transmit and receive antennas in respective order, $F_s$ is the sampling frequency, and $N$ is the \ac{ofdm} \ac{fft} size. Finally, $\vec{a}_{\text{TX}}(\cdot)\in \mathbb{C}^{N_{\text{TX}}}$ and $\vec{a}_{\text{RX}}(\cdot)\in \mathbb{C}^{N_{\text{RX}}}$ are the steering vectors, which define the phases per antenna element with respect to the array center, for given \ac{aod} and \ac{aoa}}.
Considering further the analog phased-arrays in \ac{mmw} systems,  the \ac{tx} and \ac{rx} apply beamforming weights $\vec{b}_{\text{TX}}\in \mathbb{C}^{N_{\text{TX}}}$ 
and $\vec{b}_{\text{RX}}\in \mathbb{C}^{N_{\text{RX}}}$, respectively.
The corresponding effective beamformed channel at subcarrier $n$, considered as the \emph{frequency-domain \ac{csi}}, can then be expressed as 
\begin{equation} \label{eq:Hn}
    \mathcal{H}(n) = \vec{b}_{\text{RX}}^H \boldsymbol{\mathcal{H}}(n) \vec{b}_{\text{TX}}. 
\end{equation}

\textls[-2]{In practice, besides noise and interference, the \ac{csi} estimation can suffer from inaccuracies~\cite{ferrand2020dnn} due to \ac{rf} impairments, clock and frequency offsets between the \ac{gnb} and the \ac{ue}, and imperfect timing advance information. 
Furthermore, due to signaling overhead, \ac{csi} is often reported per blocks of subcarriers, which reduces the \ac{csi} resolution in frequency. In this paper, we consider obtaining the frequency-domain \ac{csi} via \ac{5g} \ac{nr} \ac{ss}s, transmitted periodically in \ac{dl} by all \ac{gnb}s.}

\subsection{Time-Domain Multipath Measurements through \ac{mrtt} and \ac{ulaoa}}
\label{sec:3td}

In time-domain, the radio propagation channel can be modeled as a composition of individual propagation paths with path-specific propagation delay, power gain, phase shift, \ac{aod}, and \ac{aoa}, together with additional distortion and interference, among other channel effects. The antenna-element-wise channel impulse response $\mathbf{H}(\tau) \in \mathbb{C}^{N_{\text{RX}} \times N_{\text{TX}}}$ can be written as a function of propagation delay $\tau$ as \cite{chen2022joint, 3GPPchannel}
\begin{equation}
  \mathbf{H}(\tau) = \sum_{k=0}^{K-1} h_k \vec{a}_{\text{RX}}(\theta_{\text{AOA},k}) \vec{a}_{\text{TX}}^H(\theta_{\text{AOD},k}) \delta(\tau-\tau_k)
\end{equation}
where $\delta(\cdot)$ is a Dirac delta function (i.e., a unit impulse). 
Similar to the frequency-domain representation, while again assuming analog phased-arrays, the effective beamformed channel impulse response can be written as 
\begin{equation}
    H(\tau) = \vec{b}_{\text{RX}}^H \mathbf{H}(\tau) \vec{b}_{\text{TX}}. 
\end{equation}

In this paper, the measured \emph{time-domain \ac{csi}} includes the path delays $\tau_k$, the path powers $\vert h_k \vert^2$, and the \ac{gnb} side path angles $\theta_{\text{AOA},k}$. For the path delays $\tau_k$ and path powers $\vert h_k \vert^2$, the estimation procedure is assumed to exploit \ac{5g} \ac{nr} \ac{mrtt} measurements~\cite{3GPPue}, as discussed in Section \ref{sec:3basics}. Furthermore, for estimating the \ac{gnb} side path angles $\theta_{\text{AOA},k}$, it is also possible to utilize \ac{ulaoa} measurements~\cite{3GPPue} based on \ac{ul} \ac{srs} transmissions, as noted in Section \ref{sec:3basics}.

The overall data acquisition concept is illustrated in Fig. \ref{fig:dacq}, highlighting the different considered measurements. In general, within the current \ac{5g} \ac{nr} standard, the \ac{lmf} is responsible for the localization and related signaling management while the positioning calculations can be carried out either at the \ac{ue} or the network side. Moreover, reporting the \ac{mrtt} and \ac{ulaoa} measurements between the \ac{gnb} and the \ac{lmf} is supported by the so-called \ac{nr} Positioning Protocol A~\cite{3GPPpos}. \blue{Different alternative ways to arrange for labeled training data include crowdsourcing, crowdsensing, as well as utilization of synthetic data. These are discussed further in Section IV.G.}

\section{Proposed Methods}
\label{sec:4}

This section describes and introduces the novel approach of utilizing frequency-domain \ac{csi} with relative phase, while also addressing the time-domain \ac{csi} data pre-processing. In addition, the proposed architectures, hyperparameters, and training algorithm of the proposed \ac{nn} models are presented. \red{Finally, important system-level implementation alternatives and aspects are discussed.}

\vspace{-3mm}
\subsection{Frequency-Domain CSI Data Preprocessing}
\label{sec:4b}

\subsubsection{Proposed Relative Phase Approach}
\label{sec:4b1}
\ac{5g} \ac{mmw} networks operate at high carrier frequencies, at and beyond 24\,GHz, with wavelengths approaching the millimeter-scale. In mobile scenarios, utilizing absolute phase responses is highly impractical, as movement of a few millimeters in distance results in a full rotation of the phase. Furthermore, as is well-known, the frequencies and wavelengths relate through
\begin{equation} \label{eq:lambda}
    \lambda = \delta_s =  {c}/{f},
\end{equation}
where $\delta_s$ is the propagation distance between two points with equal phases,  $\lambda$ is the signal wavelength, $c$ is the speed of light, and $f$ is the signal frequency.

In this article, we consider obtaining the frequency-domain \ac{csi} in the resolution of 12 subcarriers, which refers to a bandwidth of one \ac{rb} in \ac{5g} \ac{nr}, denoted as $\Delta f_{RB}$. The \ac{csi} is interpreted at the $6^{\text{th}}$ subcarrier of each \ac{rb}, and thus the corresponding subcarrier index for the $m^{\text{th}}$ \ac{rb} observation is given as $n_{\text{RB,}m}=6+m\Delta f_{RB}$ with $m = 0,...,M-1$, where $M$ is the number of \ac{rb}s. Moreover, we propose to take advantage of \emph{differential phase measurements between neighboring \ac{rb}s}. Based on \eqref{eq:frequency_response}-\eqref{eq:Hn}, and when considering an individual propagation path, the phase difference between subcarriers is equal across the spectrum and completely determined by the \ac{tof} through the complex exponential term. {Possible phase rotations due the other terms in \eqref{eq:frequency_response} are constant over all subcarriers, and thus do not induce phase difference between the subcarriers.} Thus, \red{for an individual path with \ac{tof} of $\tau_0$},  the phase difference between two consecutive resource blocks can be given as 
\begin{equation}
\label{eq:relative_phase}
    \Delta\phi = 2\pi\tau_0\Delta f_{RB}.
\end{equation}
\red{While the above expression builds on a single propagation path, we utilize this approach in this work also in case of realistic multipath propagation. As elaborated further below, the differential phase approach allows to mitigate the effect of phase periodicity, and thus extract relevant features for the proposed \ac{nn}-based positioning.}

To this end, the linkage between the relative phase and a specific propagation distance is unambiguous only when the relative phase is within one phase cycle ($\Delta\phi < 2\pi$). A distance $d_\phi$, which inflicts the full $2\pi$ cycle of the relative phase between two neighboring \ac{rb}s, can be solved based on \eqref{eq:relative_phase} as 
\begin{equation}
\label{eq:relative_distance}
    d_\phi = \frac{c}{\Delta f_{RB}}
\end{equation}
by denoting $\tau_\phi\Delta f_{RB}=1$, where $\tau_\phi = d_\phi/c$ is the corresponding \ac{tof} resulting in a full phase cycle. By using the relative phase difference $\Delta\phi$, instead of an absolute phase, as the frequency-domain feature, the positioning performance can be significantly improved, as shown in Section \ref{sec:5}. Although the distance ambiguity issue still remains with the phase difference recurrence at every $d_\phi$~meters, it is greatly improved compared to the recurrence level with an absolute phase at every $\delta_s$~meters, as $f \gg \Delta f_{RB}$. 

\subsubsection{Frequency-Domain CSI Features and Visualization}
\label{sec:4vis}

\begin{figure*}[t!]
  \centering
  \vspace{-2mm}
  \subfloat[]{\includegraphics[width=0.28\textwidth]{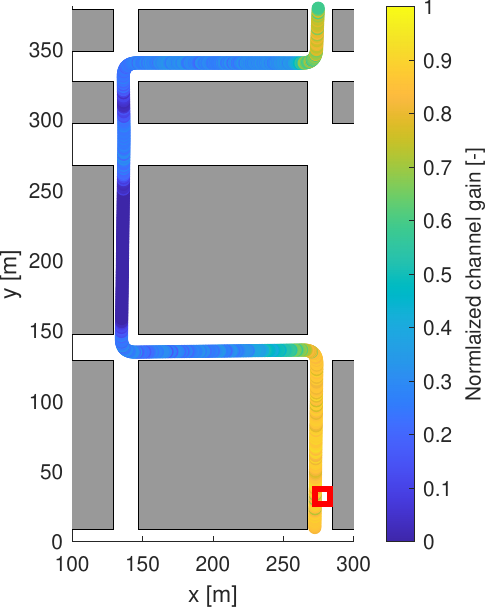}\label{fig:track}}
  \hfill
  \subfloat[]{\includegraphics[width=0.32\textwidth]{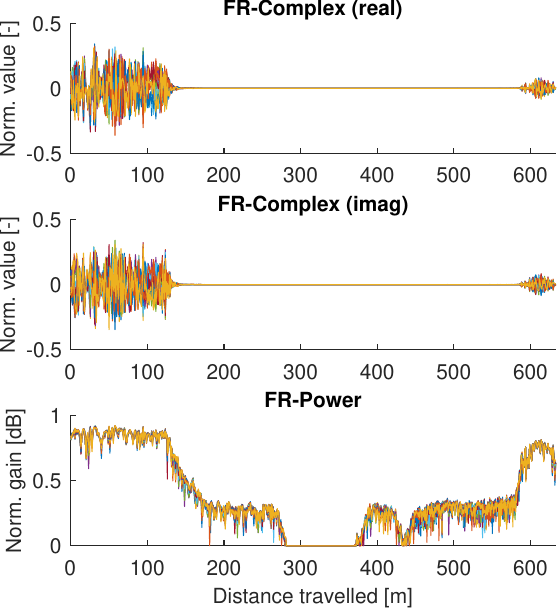}\label{fig:prep1}}
  \hfill
  \subfloat[]{\includegraphics[width=0.32\textwidth]{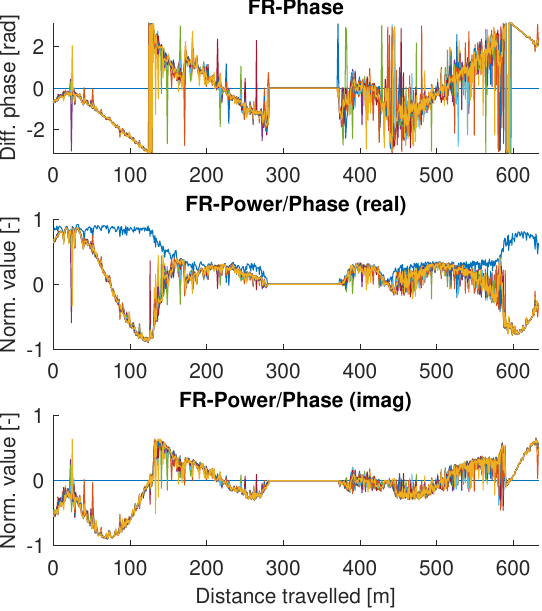}\label{fig:prep2}}
  
  \vspace{-2mm}
  \caption{\blue{An example UE track in urban environment is shown in (a) where the \ac{gnb} location is depicted with the red rectangle while the track color represents the mean normalized power across the resource blocks. The frequency-domain features \ac{fr-c} and \acs{fr-pow} as well as the proposed \acs{fr-ph} and \acs{fr-pp} are visualized in (b) and (c), respectively, along the UE track shown in (a).} \blue{In (b) and (c), different colors represent the ten different RB allocations within the full SSB transmission bandwidth.}}

  \vspace{-2mm}
\end{figure*}

To provide a short illustration, we consider a single representative user path along an urban environment, as shown in Fig.~\ref{fig:track} (for further details of the environment, refer to Section~\ref{sec:5}). Then, 
Fig.~\ref{fig:prep1} and Fig.~\ref{fig:prep2} demonstrate the utilized frequency-domain \ac{csi} feature representations along the path, including the proposed features and the features from the related literature.
Specifically, the raw, complex channel response, further {referred to as \ac{fr-c}}, is depicted in Fig.~\ref{fig:prep1}, top, and Fig.~\ref{fig:prep1}, center, which show the real and imaginary parts of \ac{fr-c} for 10 consecutive resource blocks along the path. The feature is obtained from \eqref{eq:Hn} as $\text{real}(\mathcal{H}(n_{\text{RB,}m}))$ and $\text{imag}(\mathcal{H}(n_{\text{RB,}m}))$ for $m = 1,...,10$. 

Furthermore, Fig.~\ref{fig:prep1}, bottom, depicts the channel power response, {denoted as the \ac{fr-pow}}. Such channel feature is utilized, e.g., in~\cite{wang2016csi, chen2017confi}, and can be expressed via \eqref{eq:Hn} as $10\text{log}_{10}(|\mathcal{H}(n_{\text{RB,}m})|^2)$. 
In \ac{5g} \ac{nr}, the \ac{fr-pow} corresponds to a \ac{rb}-wise \ac{rsrp} measurement, defined as the average power of the resource elements carrying the reference symbols. {As an input feature, we also re-scale \ac{fr-pow} to the normalized range of $[0,1]$}.

Then, the top graph of Fig.~\ref{fig:prep2} visualizes the proposed relative phase difference as the frequency-domain feature, {further denoted as the \ac{fr-ph}}. Specifically, building on the discussion in Section~\ref{sec:4b1}, the \ac{fr-ph} can be obtained and expressed following \eqref{eq:Hn} as 
\begin{equation}
\label{eq:relative_phase_RB}
    \Delta \phi(m) = \text{arg}(\mathcal{H}(n_{\text{RB,}m}))-\text{arg}(\mathcal{H}(n_{\text{RB,}m-1}))
\end{equation}
for \ac{rb} indices $m=1,...,M-1$.
The dependency between the signal path lengths and the relative phases, especially in \ac{los} regions, is clearly visible in the figure. Moreover, it can be seen that the feature magnitude is recurring with a path propagation distance at every $d_\phi$ meters, as derived in \eqref{eq:relative_distance}.

\subsubsection{Proposed Combined Feature}To utilize the maximum information enclosed in the measured channel responses, we further propose the so-called \ac{fr-pp} approach as the ultimate frequency-domain feature. This approach combines the \ac{fr-ph} and \ac{fr-pow} by transforming the \ac{fr-ph} to the complex unit circle, with subsequent element-wise multiplication with the re-scaled \ac{fr-pow}. This is expressed as
\begin{equation}
\label{eq:FR-phase}
    \bar{P}_\text{RB}(m)\,\text{exp}({j\Delta \phi (m)})
\end{equation}
where $\bar{P}_\text{RB}(m)$ refers to the re-scaled normalized power for RB indices $m=0,...,M-1$.
Furthermore, since the  \ac{fr-ph} has one element less than \ac{fr-pow}, we extend the \ac{fr-ph} array with an additional element for $m=0$ by defining $\Delta \phi (0) = 0$.  

The real and imaginary components of the proposed \ac{fr-pp} feature set are visualized in Fig.~\ref{fig:prep2}, center and bottom, respectively. The proposed feature allows to accommodate the advantages of both received power and relative phases in a single complex feature vector, while relaxing the distance ambiguity of the relative phase feature.

\vspace{-2mm}
\subsection{Time-Domain CSI Data Preprocessing}
\label{sec:4netf}

The time-domain \ac{csi} data utilized for localization includes propagation delays $\tau_k$, powers $\vert h_k \vert^2$, and \ac{gnb} side path angles $\theta_{\text{AOA},k}$ for the observed \ac{los} and \ac{nlos} paths. 
To this end, the measured path-wise propagation delays $\tau_k$, {referred to as \ac{p-tof},} are transformed to propagation distances by multiplying them with the speed of light. 
The path-wise \ac{aoa}s, $\theta_{\text{AOA},k}$, are obtained at the \ac{gnb} side, and transformed to directions in Cartesian coordinates in the preprocessing, to omit the zero-crossing problem with cyclic angular data. This results in a robust \ac{aoa} feature, {called \ac{p-aoa} in the following,} which is less susceptible to angular deviation and related uncertainties.
The {\acp{p-rp} $\vert h_k \vert^2$ are expressed in decibels (dBm)} to overcome the extremely low feature magnitudes in linear scale. Such feature is the time-domain equivalent of the frequency-domain \ac{fr-pow}, which accumulates all paths into the same observed frequency-domain measurement. Similar to the propagation delay feature,  the path power feature includes information on the path propagation distance, but most importantly, it also provides information on the number and type of channel interactions, such as reflections,  diffraction, or scattering, within the radio path. {In general, different combinations of the time-domain CSI data can be adopted. The aggregated path-ToF+AoA and path-ToF+RP+AoA features, proposed in this work, are the most powerful ones, as shown through the numerical results.}

\vspace{-3mm}
\subsection{NN Model Architectures and Hyperparameters}
\label{sec:4nn}
 
Among the various alternative data-aided approaches, we restrict ourselves to \ac{nn} models in this work, which currently dominate the \ac{ml} area due to their performance, scalability, generalization properties, and dynamic architecture options~\cite{goodfellow2016deep}.

\subsubsection{\red{Activation Function}}
\label{sec:4nnbasics}

In this work, we utilize the \ac{gelu}~\cite{hendrycks2016gaussian} as the non-linear activation function. Its main advantages over traditional \ac{relu} include \red{resistance to a ``dying ReLU'' problem~\cite{lu2019dying},} differentiability at all values while having also been shown to offer improved performance already in a number of applications such as natural language processing~\cite{hendrycks2016gaussian}. 
It can be defined as $\text{GELU}(x) =x\Phi(x)$~\cite{wang2020transformer, xiao2022image} 
where $\Phi(x)$ is the cumulative distribution function of the standard normal distribution. The function can also be approximated for faster processing as
\begin{equation} \label{geluappr}
    \text{GELU}(x) = \frac{x}{2} \tanh\left(\sqrt{\frac{2}{\pi}(x+C x ^3)}\right),
\end{equation}
where $C = 0.044715$.
Compared to \ac{relu}, the higher complexity of \ac{gelu} is compensated by the faster convergence of the model\red{, as well as the corresponding improved positioning performance, based on our complementary experiments}.

\subsubsection{Utilized NN Architectures} 
\label{sec:4nnmodels}

As the functional \ac{nn} layers, we utilize in this work both densely connected layers and \ac{lstm}~\cite{hochreiter1997long} layers, the later being used only in the sequence-based implementation of models.
Importantly, the \ac{lstm} layer is a recurrent-based layer capable of preserving long-term and short-term trends within the data.

\begin{figure}[t!]
  \centering
  \subfloat[]{\includegraphics[width=0.42\textwidth
  ]{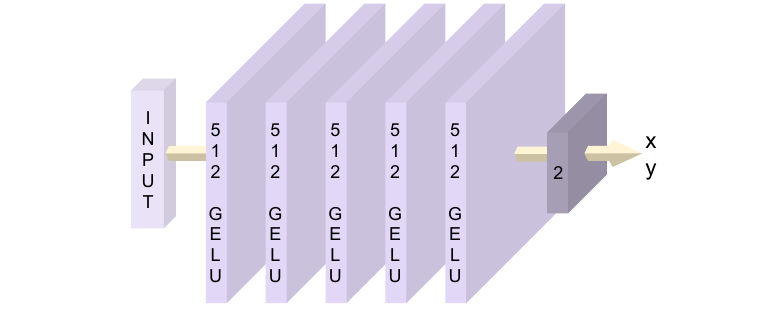}\label{fig:nn1}}
  \\
  \vspace{-2mm}
  \subfloat[]{\includegraphics[width=0.42\textwidth,
  ]{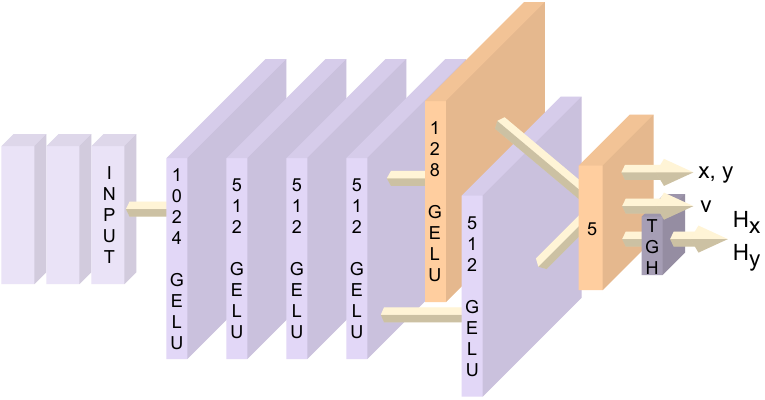}\label{fig:nn2}}
  \vspace{-1mm}
  \caption{Architecture and hyperparameters of the densely connected \ac{nn}, in (a), and of the sequence processing \ac{nn}, in (b). Each layer is specified by the number of neurons and an activation function.}
  \vspace{-2mm}
\end{figure}

The architecture of the densely-connected model is depicted in Fig.~\ref{fig:nn1}. It consists of 5 densely connected layers with \ac{gelu} activation functions and a single densely connected layer with linear activation and 2 neurons as the output, estimating the \ac{ue} position. 
The architecture of the sequence processing capable model is, in turn, shown in Fig.~\ref{fig:nn2}. It consists of 5 densely connected layers after the input, with a single \ac{lstm} layer connected in parallel with the 5$^{th}$ dense layer. The concatenated output of these layers is then fed to an \ac{lstm} layer with 5 neurons at the output with linear activation. \red{Specifically, the densely connected layers serve as instantaneous feature extractors, while the intermediate \ac{lstm} layer learns the temporal features. Due to the considered parallel architecture, the last functional layer has access to both instantaneous and temporal features.
}  The resulting output is then divided into a positioning output with $2$ variables, velocity output with a single variable, and a heading output with an additional $\text{tanh}(\cdot)$ activation and $2$ variables.

\red{
\subsubsection{Data Structures, Normalization and Training}
In general, the input dimensions vary based on the selected features and deployment scenario. For frequency domain features, and when considering the evaluation scenario described in Section IV containing $3$ \acp{gnb}, $16$ beams, and $10$ \acp{rb}, the input size is either $480$ for \ac{fr-pow} and \ac{fr-ph} or $960$ for \ac{fr-pp} and \ac{fr-c}. When considering the time-domain CSI in the same scenario, the individual \ac{p-tof}, \ac{p-aoa}, and \ac{p-rp} features have each an input size of $15$, the combined path-ToF+AoA and path-RP+AoA 
features have $30$ inputs, and finally the aggregated path-ToF+RP+AoA feature has an input size of $45$. Some of the features are also normalized prior to the training, as demonstrated already along Fig.~\ref{fig:prep1} and Fig.~\ref{fig:prep2}.} {Specifically, all power-related quantities as well as path-wise \ac{tof} measurements, when first converted to pseudo-ranges, are all normalized between $[0,1]$ within the overall sets of available measurements.} {Finally, all angle and phase quantities are, by design, within $[-\pi,\pi]$.} \blue{We emphasize that each different feature scenario and combination corresponds to an individual \ac{nn}, trained and deployed on its own. The vast set of numerical results, provided in Section IV, provides the corresponding mutual performance comparisons.}

All considered \ac{nn} models are trained using the Adam optimizer~\cite{kingma2014adam} with learning rates of $0.001$ for the first $200$ epochs, and then an early-stopping mechanism based on validation performance for additional $500$ epochs, while iteratively reducing the learning rate to $0.0005$ and $0.0001$ after each stop. The lowered learning rates ensure a fine-tuned performance with a small number of epochs. The mean squared error (MSE) loss was selected for each output, and for the sequence-based \ac{nn} model, the loss weights were selected as $0.8$, $0.1$, and $0.1$ for positioning loss, velocity loss, and heading loss, respectively. \red{Furthermore, stemming from the deployment area of around $550\times370~$m$^2$ (see Fig.~5), the position labels are reduced by a factor of $300$ to accelerate the training.}

\vspace{-2mm}
\red{
\subsection{System-Level Implementation Alternatives and Aspects}

In general, there are alternative ways to organize and implement the use of the CSI measurements and data for \ac{nn} training and actual online inference processing for localization. These are discussed below, \blue{in relation to the proposed methods and the data acquisition visualized in Fig.~1}, while noting also the important role of \ac{ue} \ac{rrc} state. 

To this end, the time-domain CSI data, i.e., the \ac{mrtt}-based \ac{tof} measurements and the SRS-based \ac{ulaoa} measurements, are by definition obtained at the network side. Thus, in this case, it is natural to also perform both the model training as well as the localization inference processing at the network side. Consequently, there is no need for additional signaling or feedback, and all training data from different \acp{ue} is inherently gathered together for training the model. Importantly, since \ac{mrtt} and \ac{ulaoa} require scheduled \ac{srs} and \ac{prs} transmissions, time-domain measurements are only available in the \emph{connected mode} when it comes to the \ac{ue} \ac{rrc} state. 

Frequency-domain \ac{rsrp} and other \ac{csi} measurements are collected from periodic and always available \ac{ssb} transmissions at the \ac{ue} side, thus enabling utilization of efficient data crowdsourcing methods. Despite a possible technical capability to perform training at the \ac{ue},  assuming individually trained models at different \acp{ue} can be considered unrealistic. Therefore, \acp{ue} are expected to periodically share such measurement data with the network for \ac{nn} training, for example, through \ac{mdt} messaging in the form of raw measurements and location tags, or alternatively as locally pre-trained models following the principle of federated learning (FL). Interestingly, unlike with \ac{mrtt} and \ac{ulaoa}, the \ac{dl} frequency-domain \ac{csi} and \ac{rsrp} measurements can be collected and obtained also in the \ac{rrc} \emph{idle mode} as part of standard mobility management procedures. This can be considered a great asset enabling continuous data collection and localization with very low power consumption. Furthermore, assuming a pre-shared model from the network for the final inference phase, the \ac{ue} can perform localization independently without supplementary signaling with the \ac{gnb}.

}
\vspace{-3mm}
\section{Evaluation Environment and Results}
\label{sec:5}
\subsection{Evaluation Scenario and Assumptions} \label{sec:5.2}
The evaluation environment builds on ray-tracing-based channel measurements utilizing Wireless Insite\textregistered software~\cite{wirelessInsite}. 
We employ the map-based METIS Madrid grid~\cite{madridgrid}, recognized as the relevant urban scenario by 3GPP in 5G NR specifications~\cite{3GPPchannel}. The Madrid grid layout introduces generally a rich radio propagation environment with different street widths and open areas, empowering generalization and scalability. 

\begin{figure}[t!]
    \centering
    \vspace{-0mm}
    \includegraphics[width=1.0\columnwidth]{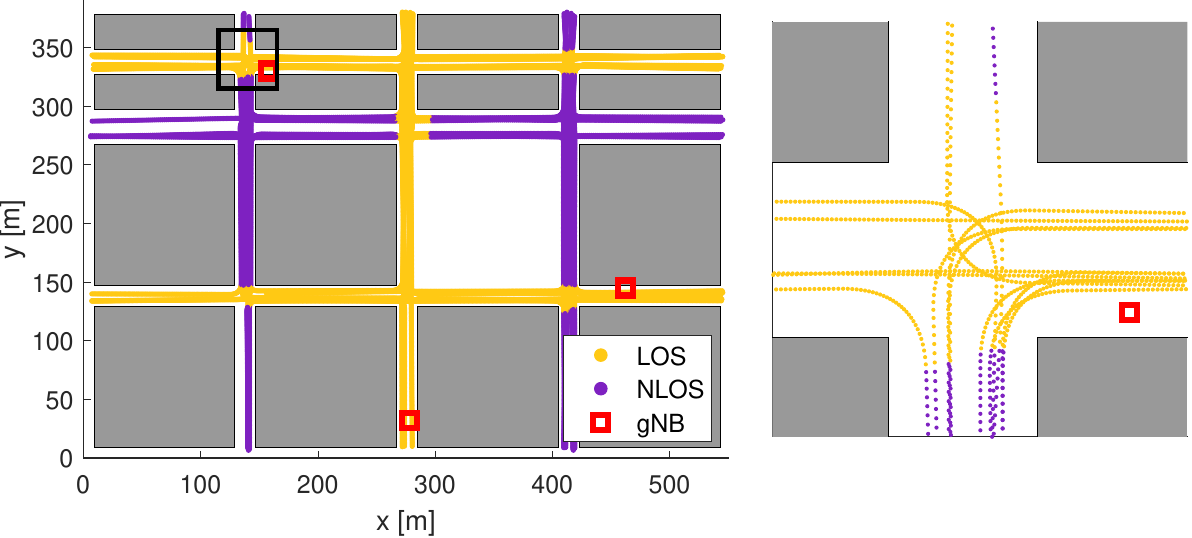}
    \captionsetup{belowskip=0pt}
    \linespread{1.0}
    \vspace{-5mm}
    \caption{Illustration of the METIS Madrid map-based deployment and evaluation scenario with 40 simulated \ac{ue} tracks while distinguishing the \ac{los} and \ac{nlos} regions. {Detailed example paths on one crossing are shown on the right.}}
    \vspace{-4mm}
    \label{fig:deploy}
\end{figure}

The simulated urban scenario illustrated in Fig.~\ref{fig:deploy} contains three \ac{5g} \ac{mmw} \ac{gnb}s operating at $28$\,GHz, such that clear \ac{nlos} regions also exist along the streets. Each \ac{gnb} is equipped with a uniform cylindrical antenna array with $4$ elevated layers, each with $16$ antenna elements placed at $5$\,m height. The beam configuration includes 16 beams with uniformly separated azimuth angles and a common down-tilted elevation angle fixed at $10$\,deg. 
The \ac{aod} and \ac{tof} measurements are obtained based on the corresponding characteristics of the radio propagation path with the highest received power, building on the signals and measurement procedures described in Section~\ref{sec:3}. The obtained \ac{aod} and \ac{tof} measurements are exposed to substantial measurement errors, as discussed further in Section~\ref{sec:4unc}. The beam-wise frequency-domain \ac{csi} measurements are obtained from \ac{ssb} transmissions as an average of received subcarrier powers per RB, with $120$\,kHz subcarrier spacing. Measurements with path-loss higher than 160\,dB are not considered, while in general the environment shown in Fig.~\ref{fig:deploy} possesses large areas and street segments with severe multi-bounce phenomena.

The combined time-domain and frequency-domain dataset consists of $40$ vehicle-like user tracks, where the \ac{ue} collects measurements at $100$\,ms intervals. The \ac{ue} locations are initialized with random locations along the streets, and the \ac{ue}s move within the area by considering an equal probability to advance in any direction at intersections. The \ac{ue} velocity varies between $20$\,km/h and $60$\,km/h depending on the present street and possible proximity of intersections while when approaching an intersection, the \ac{ue} decelerates at $3$\,m/s\textsuperscript{2} until reaching a fixed velocity of $20$\,km/h for smooth turning. After the turn, the \ac{ue} accelerates at $2$\,m/s\textsuperscript{2} until reaching a street-specific speed limit. The speed limit is generally defined as $40$\,km/h, apart from the top horizontal street which has the speed limit of $20$\,km/h (see Fig.~\ref{fig:deploy}) and the wider street below the pedestrian street having a limit of $60$\,km/h. 
The exact \ac{ue} trajectories and associated measurement locations are different for each simulated user track.
As this work is heavily focused on \ac{nlos} positioning, Fig.~\ref{fig:deploy} visualizes the simulated tracks with the \ac{los}/\ac{nlos} indication at each sampled location. \blue{We note that in order to efficiently track moving \acp{ue} with varying velocities through the sequence processing models, a sufficiently rich training dataset is needed with representative velocity statistics.}

\textls[-1]{The available $40$ user tracks are distributed into 32 \ac{ue} traces for training, 4 for validation, and the remaining 4 for the actual testing. The validation and testing paths are carefully selected, 
to avoid any area-specific bias in the evaluation. Furthermore, as the work focuses on the \ac{nlos} positioning performance, we validated the consistency of the \ac{los}/\ac{nlos} split across the datasets. The distribution of the samples in the individual datasets based on the number of \ac{los} \ac{gnb}s is consistent with approx. $35\%$ \ac{nlos} samples, $60\%$ of samples having a single \ac{los} \ac{gnb}, and only $5\%$ samples having $2$ \ac{gnb}s in \ac{los}. The distribution suggests that the traditional model-based solutions, such as trilateration, are not applicable in the considered scenario. In total, there are $25\,181$ samples in the dataset.}

\vspace{-3mm}
\subsection{Network Data Uncertainties} 
\label{sec:4unc}
In this work, 
we take into account the important practical aspect of uncertainties in the measurements and thereon in the corresponding features. To this end, the frequency-domain \ac{csi} is impaired in its \ac{fr-c} representation with complex \ac{awgn} samples with magnitude equal to $30\%$ of the corresponding channel estimate's \ac{rms} magnitude. Such represents large practical measurement uncertainties. The other related features such as the  \ac{fr-pp} are impaired correspondingly, through the transformations from the impaired \ac{fr-c} to amplitude/power and phase domains.

To impair the time-domain features, we impose impairments separately to \ac{p-tof}, \ac{p-rp} and \ac{p-aoa} quantities. 
The \ac{p-tof} feature uncertainty is an \ac{awgn} with \ac{std} equal to $10$\,m. 
We consider the constant uncertainty scale regardless of the \ac{tof} magnitude, as the measurement errors are mostly resulting from hardware inaccuracies and timing offsets in the \ac{ue}s and \ac{gnb}s. 
Furthermore, we impair the \ac{p-rp} feature with an \ac{awgn} with $2$\,dB \ac{std}, which corresponds to the maximum impairment of $\pm6$\,dB range with $99.7\%$ certainty, defined by 3GPP as the required absolute measurement accuracy for \ac{ss}-\ac{rsrp}~\cite{3GPPrrm}.
The \ac{p-aoa} is, in turn, impaired with discretized accuracy of $22.5^\circ$ ($360^\circ/16$ beams), rather than with a randomized value, to incorporate the \ac{gnb} limitations in accurately determining the \ac{aoa}. 

\red{Finally, as reviewed in the Introduction, a large majority of the state-of-the-art works, such as \cite{8823059, gao2022towards, butt2020rf, chen2017confi, ferrand2020dnn, wang2016csi, gonultacs2021csi, klus2021neural, klus2021transfer}, utilize channel amplitude or power response, or even the integrated received power, as the positioning feature. Hence, in the following, the results with \ac{fr-pow} feature represent essentially the state-of-the-art reference approach when it comes to the frequency-domain features. In the time-domain feature case, the use of the individual dominant path features has been considered in \cite{10086654, 9832776, 7984759, lynch2020localisation, hajiakhondi2021bluetooth, chen2022joint}, thus serving as the main reference schemes. Additionally, the state-of-the-art schemes build commonly on snap-shot NNs without harnessing the temporal correlation.}

\blue{Data and codes are openly available at \url{https://doi.org/10.5281/zenodo.12204893}.}

\vspace{-2mm}
\subsection{Numerical Results with Dense Snap-Shot NNs} \label{sec:5.3}
We next provide and analyze the results obtained with dense \ac{nn} based \ac{ml} models while considering both the frequency-domain and time-domain features as well as the impacts of the feature density or granularity in the two considered domains. To establish an understanding on the baseline or reference performance, we start with the results under perfect measurements (no uncertainties), while then show also the performance under practical measurement uncertainties.

\subsubsection{Results with Frequency-Domain Features} \label{sec:5.3freq}

First, we analyze and compare the different frequency-domain \ac{csi} features introduced in Section~\ref{sec:4b} and their positioning capabilities with a densely connected snap-shot \ac{nn} with 5 hidden layers. We thus split all the user tracks into individual samples and compare the performance without considering the temporal dependencies within sequences or additional uncertainties, to focus on the quality of the features themselves.

\begin{figure*}[t!]
  \centering
  \vspace{-2mm}
  \subfloat[]{\includegraphics[width=0.4\textwidth]{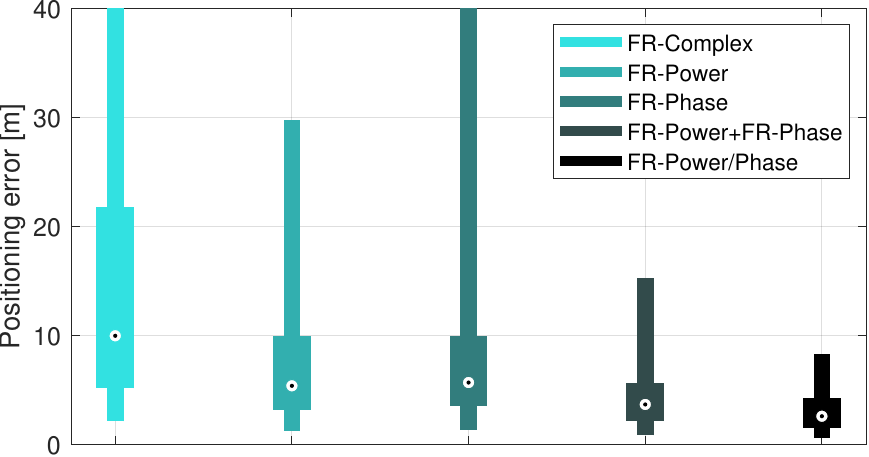}\label{fig:boxplot_features}}
  \hspace{2.0cm}
  \subfloat[]{\includegraphics[width=0.4\textwidth]{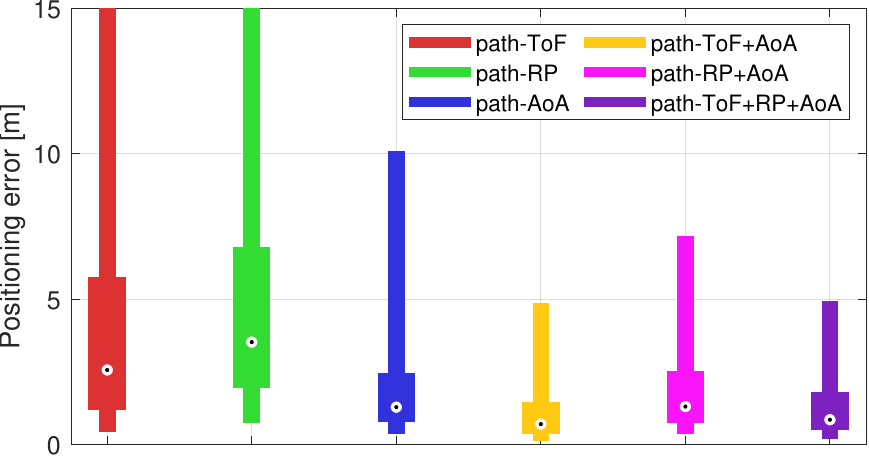}\label{fig:spawc_boxplot_features}}
  \hspace{0.0cm}
  \vspace{-2mm}
  \caption{Distributions of positioning errors on the testing data when evaluating (a) different frequency-domain features, and (b) different time-domain features, with dense snap-shot \ac{nn} and with no measurement uncertainties.}
\end{figure*}

Fig.~\ref{fig:boxplot_features} visualizes the distributions of the positioning errors on the testing dataset for each feature. Each boxplot marks the median (center) as well as the first and third quartiles ($25^{th}$ and $75^{th}$ percentiles) encapsulated in the box, while the whiskers mark the values of $5^{th}$ and $95^{th}$ percentiles.
The results show that the proposed \ac{fr-pp} feature representation enables the most efficient training in terms of positioning error and that considering the \ac{fr-c} features as the input provides the poorest performance. The \ac{fr-pow} and \ac{fr-ph} features achieve comparable median performance, but in terms of outliers, \ac{fr-pow} performs better. The $95^{th}$ percentiles, referring essentially to the presence of outliers, of \ac{fr-c} and \ac{fr-ph} are significantly higher than those of the remaining methods, as shown quantitatively in Table~\ref{tab:results_features}. Furthermore, the feature combination denoted as \ac{fr-pow}+\ac{fr-ph} represents the simple concatenation of the corresponding individual features. The numerical results show that the positioning performance is improved when compared to the individual features, but the novel \ac{fr-pp} feature -- utilizing the same, yet pre-processed inputs -- provides superior performance. The table high-lights in bold the best performance numbers in different cases. 

\vspace{-1mm}
We next further investigate the impact of the feature representation by considering the \ac{los} and \ac{nlos} data separately, with the results being shown in Table~\ref{tab:results_features}. We can observe that the proposed \ac{fr-pp} feature representation achieves the lowest positioning errors by a considerable margin, when compared to the other methods ($3.35$\,m and $4.70$\,m mean positioning error in \ac{los}/\ac{nlos}, respectively) addressed earlier in the literature. By comparing the performance in \ac{los} and \ac{nlos} scenarios, we can observe some increase in the error in \ac{nlos}, however, the exact impact is clearly feature-dependent. Furthermore, when considering the $95^{th}$ percentiles of the error distributions, we can observe that the errors related to the \ac{fr-c} and \ac{fr-ph} features are drastically increased, in both \ac{los} and \ac{nlos} scenarios, while \ac{fr-pow}+\ac{fr-ph} and \ac{fr-pp} features sustain a relatively stable performance across the majority of the testing samples.

\begin{table*}[t]
\caption{\textsc{Baseline performance results: frequency-domain features, dense snapshot NN, no uncertainties}}
\label{tab:results_features}
    \centering
    \tabcolsep 1.5pt
    \begin{tabular}{lSSSSSSSSSS}
    \toprule
    {\textbf{Feature}} & \multicolumn{2}{c}{\textbf{\ac{fr-c}}} & \multicolumn{2}{c}{\textbf{\ac{fr-pow}}} & \multicolumn{2}{c}{\textbf{\ac{fr-ph}}} & 
    \multicolumn{2}{c}{\textbf{\ac{fr-pow}+\ac{fr-ph}}} & \multicolumn{2}{c}{\textbf{\ac{fr-pp}}}\\ 
    \cmidrule(rl){2-3} 
    \cmidrule(rl){4-5}  
    \cmidrule(rl){6-7}
    \cmidrule(rl){8-9}
    \cmidrule(rl){10-11}
    {Error [\si{\meter}]} & {LoS} & {NLoS} & {LoS} & {NLoS} & {LoS} & {NLoS} & {LoS} & {NLoS} & {LoS} & {NLoS}\\
    \midrule
{Median} 	&	7.88	&	14.01	&	5.57	&	5.01	&	5.26	&	6.97	&	3.35	&	4.29	&	\textbf{2.42}	&	\textbf{2.84}	\\
{Mean} 	&	21.27	&	25.63	&	9.69	&	8.89	&	18.01	&	27.99	&	6.53	&	8.67	&	\textbf{3.35}	&	\textbf{4.70}	\\
{$80^{th}$ pc}	&	21.70	&	33.73	&	12.36	&	11.11	&	9.33	&	17.67	&	5.46	&	8.57	&	\textbf{4.45}	&	\textbf{5.18}	\\
{$95^{th}$ pc}	&	106.37	&	92.05	&	30.69	&	27.53	&	87.97	&	166.66	&	10.47	&	21.43	&	\textbf{6.71}	&	\textbf{10.97}	\\

\bottomrule

\end{tabular}

\end{table*}

\red{Overall, the obtained results clearly show and demonstrate that utilizing the novel \ac{fr-pp} feature offers the best performance by a large margin, clearly outperforming the earlier state-of-the-art in the field of frequency-domain features.} 
Thus, in the further frequency-domain feature related evaluations, we consider only the \ac{fr-pp} feature representation. 

\subsubsection{Results with Time-Domain Features} \label{sec:5.3time}
Next, we evaluate the positioning capabilities and performance when utilizing the different time-domain features (\ac{p-tof}, \ac{p-rp}, and \ac{p-aoa}) as the input data. 
We also evaluate the combination of the features, while the model can consider up to $5$ dominant multipath components above the 160\,dB path-loss threshold. 

Fig.~\ref{fig:spawc_boxplot_features} visualizes the achieved positioning results, showing that the proposed combinations of \ac{p-tof} and AoA or \ac{p-tof}, RP and AoA are the two best performing aggregate features. The results also suggest that the \ac{p-rp} measurement provides less relevant information to the model than the \ac{p-tof}, which the model can directly interpret as normalized pseudo-range measurement. \red{This can be seen by comparing the individual features (path-ToF vs. path-RP), as well as the cases where they are combined with \ac{p-aoa}.}

\begin{table*}[t]
\caption{\textsc{Baseline performance results: time-domain features, dense snapshot NN, no uncertainties}}
\label{tab:results_features_spawc}
    \centering
    \tabcolsep 1.5pt
    \begin{tabular}{lSSSSSSSSSSSS}
    \toprule
    {\textbf{Feature}} & \multicolumn{2}{c}{\textbf{path-ToF}} & \multicolumn{2}{c}{\textbf{path-RP}} & \multicolumn{2}{c}{\textbf{path-AoA}} & \multicolumn{2}{c}{\textbf{path-ToF+AoA}} & \multicolumn{2}{c}{\textbf{path-RP+AoA}} & \multicolumn{2}{c}{\textbf{path-ToF+RP+AoA}}\\ 
    \cmidrule(rl){2-3} 
    \cmidrule(rl){4-5}  
    \cmidrule(rl){6-7}
    \cmidrule(rl){8-9}
    \cmidrule(rl){10-11}
    \cmidrule(rl){12-13}
    {Error [\si{\meter}]} & {LoS} & {NLoS} & {LoS} & {NLoS} & {LoS} & {NLoS} & {LoS} & {NLoS} & {LoS} & {NLoS} & {LoS} & {NLoS}\\
    \midrule
{Median} 	&	2.74	&	2.19	&	2.96	&	4.49	&	1.10	&	1.67	&	\textbf{0.54}	&	\textbf{1.10}	& 1.09	&	1.73	&		0.71	&	1.21	\\
{Mean} 	&	11.08	&	11.14	&	10.17	&	13.50	&	2.79	&	5.03	&   \textbf{1.23}	&   3.61	&	1.91	&	4.29	&		1.32	&	\textbf{2.63}	\\
{$80^{th}$ pc}	&	8.59	&	6.05	&	6.35	&	12.51	&	2.57	&	3.96	&   \textbf{1.35}	&   \textbf{2.33}	&	2.45	&	4.16	&		1.67	&	2.71	\\
{$95^{th}$ pc}	&	45.26	&	40.84	&	29.91	&	53.60	&	8.48	&	14.92	&   3.77    &   6.86	&	5.31	&	14.88	&		\textbf{3.44}	&	\textbf{6.73}	\\

\bottomrule

\end{tabular}

\end{table*}

The impacts of the features as well as the standalone performance in  \ac{los}/\ac{nlos} are summarized in Table~\ref{tab:results_features_spawc}, while also highlighting the best-performing features in each scenario. The table shows that the combination of all features (path-ToF+RP+AoA) together with path-ToF+AoA offer the best results across all statistics. 
\red{The corresponding performance of path-RP+AoA lags already behind. When evaluating the individual features, the \ac{p-aoa} provides high-accuracy positioning capabilities with less than $2$\,m median positioning error in \ac{nlos}, as it can effectively capture the propagation patterns within the given deployment. The \ac{p-rp} and \ac{p-tof} provide, in turn, significantly poorer performance as individual features, especially when considering the higher percentile errors. These results thus clearly prove the value of the directional measurements.} 
Additionally, when compared to the results presented in Table~\ref{tab:results_features}, relative performance improvement can be observed, which we credit to stronger interpretability of time-domain measurements as model inputs 
compared to the frequency-domain \ac{csi} features. Notably, \red{meter-scale positioning accuracy can be reached through the time-domain features also in \ac{nlos}.} 

\subsubsection{Impact of Feature Granularity}  \label{sec:5.3dens}
\textls[-1]{Next, \blue{we assess and compare the performance of the snap-shot \ac{nn} model 
while varying the granularity or sparsity of the input measurements.} 
We again separate the testing dataset into \ac{los} and \ac{nlos} parts, 
and first evaluate the frequency-domain data as \ac{rbl} features, and their mean values across the \ac{rb}s as the \emph{\ac{bwl} features}.
The \ac{rbl} features are obtained per-\ac{rb}, which contains $12$ subcarriers with a subcarrier spacing of $120$\,kHz, thus representing a bandwidth of $1.44$\,MHz per measurement. The \ac{bwl} feature considers the $14.4$\,MHz bandwidth across $10$ \acp{rb}.
Technically, the full feature (\ac{rbl}) dataset consists of $960$ features per sample (real and imaginary part, $10$~\ac{rb}s, $16$~beams, $3$~\ac{gnb}s), while the \ac{bwl} feature dataset contains only $96$ features per sample ($2\times16\times3$). The purpose is to investigate and understand whether the more sparse feature representations, which enable simpler and thus faster models, are capable of achieving competitive performance to the non-sparse data or full features.} 

\begin{figure*}[t!]
  \centering
  \vspace{-2mm}
  \subfloat[]{\includegraphics[width=0.42\textwidth]{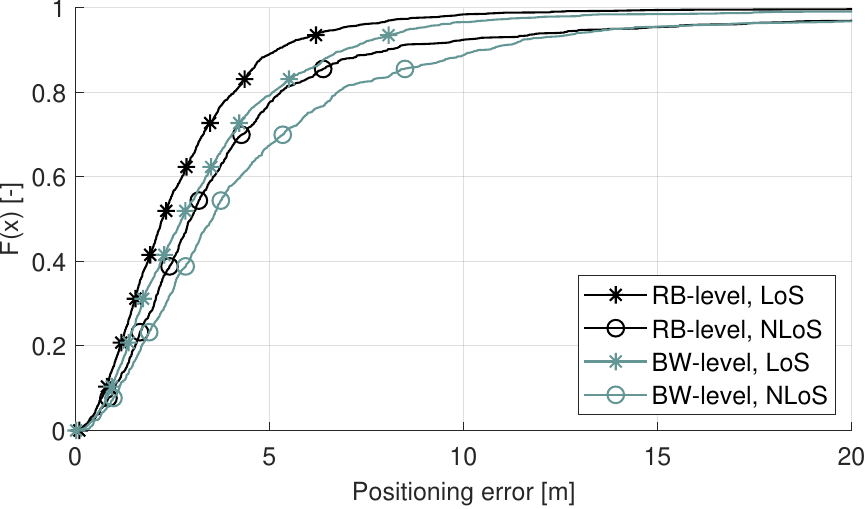}\label{fig:mean}}
  \hspace{2.0cm}
  \subfloat[]{\includegraphics[width=0.42\textwidth]{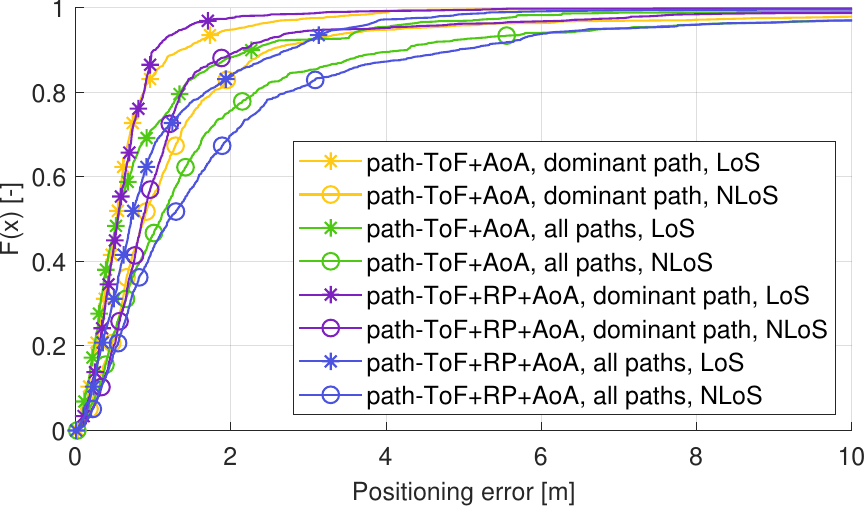}\label{fig:spawc_sparse}}
  \hspace{0.cm}
  \vspace{-2mm}
  \caption{\Acp{ecdf} of the positioning errors on the test data, (a) when evaluating \ac{rbl} and \ac{bwl} \ac{fr-pp} data, and (b) when evaluating path-ToF+AoA and path-ToF+RP+AoA features with either dominant path only or with all detected paths. The figures visualize the results of \ac{los} and \ac{nlos} regions separately using a dense snap-shot \ac{nn} model with no measurement uncertainties.}
\end{figure*}

Fig.~\ref{fig:mean} visualizes the \acp{ecdf} of the positioning errors, in the \ac{los} and \ac{nlos} regions, when utilizing the \ac{rbl} and \ac{bwl} data as the \ac{fr-pp} features. We can observe that in the \ac{los} scenario, the \ac{bwl} features provide somewhat reduced performance compared to the full \ac{rbl} features. In the \ac{nlos} scenario, the \ac{rbl} features as the input again outperform the \ac{bwl} features.
In general, particularly in \ac{nlos} where the channel geometry is more complex, the wider array of inputs can support the model in extracting more relevant positioning information thus leading to an improved performance. On the other hand, one can also conclude that the \ac{bwl} features provide a well-working solution with large reductions in the model complexity.   

Similarly, we next evaluate the impact of the time-domain feature granularity. Earlier, we already concluded that the path-ToF+AoA and path-ToF+RP+AoA features provide the best performance, thus these features are utilized also here. In the following, we distinguish and compare between utilizing only the dominant multipath component (in terms of power) and all available multipath components as the input features.

The achieved performance results are depicted in Fig~\ref{fig:spawc_sparse}, from which we can draw the following observations. The model performance actually improves when only the dominant multipath component is used as the input. This applies to both \ac{los} and \ac{nlos} regions, and the impact is particularly clear when the path-ToF+RP+AoA feature case is considered. Additionally, Fig.~\ref{fig:spawc_sparse} shows that especially the outlier performance (the highest $10\%$ of errors) is clearly improved, particularly in the \ac{los} regions. 
Overall, the results in Fig~\ref{fig:spawc_sparse} demonstrate that very high positioning performance can be achieved also in \ac{nlos} given that proper path features are utilized. 

\subsection{Numerical Results with Temporal Sequence Models} 
\label{sec:5.3all}
In this section, we evaluate the selected frequency-domain and time-domain \ac{csi} features,  namely \ac{rbl} \ac{fr-pp}, path-ToF+{RP}+AoA with the dominant component, and path-ToF+AoA with the dominant component, in the spirit of vehicular user tracking. For this purpose, we utilize the novel temporal sequence-processing \ac{nn} model proposed and described in Section~\ref{sec:4nn}, while estimating the \ac{ue} position, \ac{ue} velocity and \ac{ue} heading simultaneously. Additionally, and importantly, we now also consider the realistic uncertainties in all considered measurements and the corresponding features, as introduced in Section~\ref{sec:4unc}.
 
\begin{figure}[t]
    \centering
    \includegraphics[width=0.9\columnwidth]{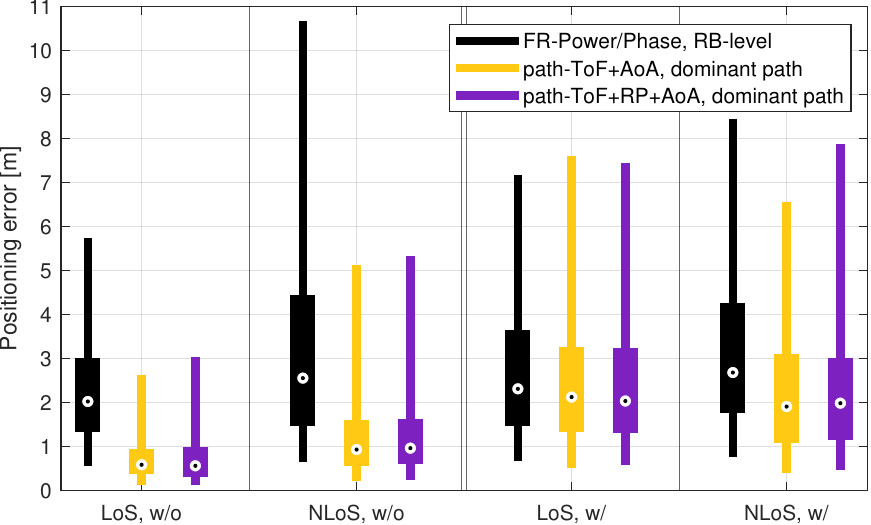}
    \captionsetup{belowskip=0pt}
    \linespread{1.0}
    \caption{Distributions of the positioning errors with sequence-processing \ac{nn} model without (w/o) and with (w/) feature uncertainties along the \ac{los} and \ac{nlos} regions. The box center denotes the median value, box edges the $25^{th}$ and $75^{th}$ percentiles, while the whiskers mark the $5^{th}$ and $95^{th}$ percentiles.}
    \label{fig:boxplot}
\end{figure}

Fig.~\ref{fig:boxplot} provides and visualizes the sequence-based model performance for the different considered features while also explicitly comparing the cases without and with measurement uncertainties, hence providing valuable insight into the \ac{nn} model generalization capabilities. In the uncertainty-free scenario, time-domain features clearly outperform the frequency-domain ones, similar to the earlier results with snap-shot models. In the practical case where the uncertainties are present in the data, the performance with the time-domain path-ToF+AoA and path-ToF+RP+AoA features deteriorates to a certain extent, while the performance gap between the \ac{los} and \ac{nlos} scenarios also interestingly disappears. The frequency-domain \ac{fr-pp} features provide essentially the same distributions for the \ac{los} and \ac{nlos} positioning errors, while even outperforming the uncertainty-free model in \ac{nlos}. These findings high-light the generalization properties of the \ac{nn} model, as the \ac{nlos} scenario itself is a source of additional uncertainties, as can be seen in Fig.~\ref{fig:prep2} already. 
To this end, Fig.~\ref{fig:boxplot} demonstrates that the increased amount of uncertainty in the data does not necessarily lead to the degradation of performance, given that the model is trained on sufficient amount of data containing such uncertainties.

To further study the achievable performance of the proposed sequence-processing model and the channel features, the positioning, speed, and heading estimation performance is next assessed and shown. 
We also compare the proposed sequence model's performance against the selected benchmark solutions, namely the \ac{nn} model from our initial work in \cite{klus2022machine} (denoted in the continuation as SPAWC benchmark), as well as an \ac{ekf}-based robust Bayesian tracking algorithm. To this end, the SPAWC model's \ac{lstm} architecture, parameters, and training setup are as described in \cite{klus2022machine}, whereas the training and testing data are naturally the same as for all other methods described in this article such that comparative results and fair comparisons can be obtained. The SPAWC benchmark utilizes path-ToF+AoA time-domain features as the inputs.

\textls[-1]{The considered \ac{ekf} benchmark, in turn, utilizes \ac{tof} and \ac{aoa} measurements from each \ac{gnb} with \ac{los} connection, while assuming ideal \ac{los} detection such that the \ac{ekf} performance is the best possible in all \ac{los} locations. In the absence of any \ac{los} link, only the prediction stage of the \ac{ekf} is conducted. The implementation of the used state-transition model and observation model, together with the related Jacobians and process covariance matrix, follows the descriptions given in \cite{koivisto2017joint}. Consequently, the used \ac{ekf} state vector comprises the \ac{ue} position in x- and y-coordinates and the \ac{ue} speed in x-y directions. For each track, the \ac{ue} state vector is initialized with a perfect state vector estimate in order to provide the best available performance for the \ac{ekf} benchmark results. However, the performance is evaluated only after the first \ac{los} link is obtained, so that the unrealistic prediction during \ac{nlos} condition with the perfect initialization is avoided. After a brief optimization of the \ac{ekf} parameters, the \ac{std} of \ac{tof} and \ac{aoa}, included in a diagonal measurement covariance matrix, are defined as 50\,ns and 15\,deg, respectively. In addition, the \ac{std} of the velocity noise, used in the process covariance matrix, as defined in \cite{koivisto2017joint}, is set as 8\,m/s.}

\begin{figure*}[t!]
  \centering
  \vspace{-2mm}
  \subfloat[]{\includegraphics[width=0.42\textwidth]{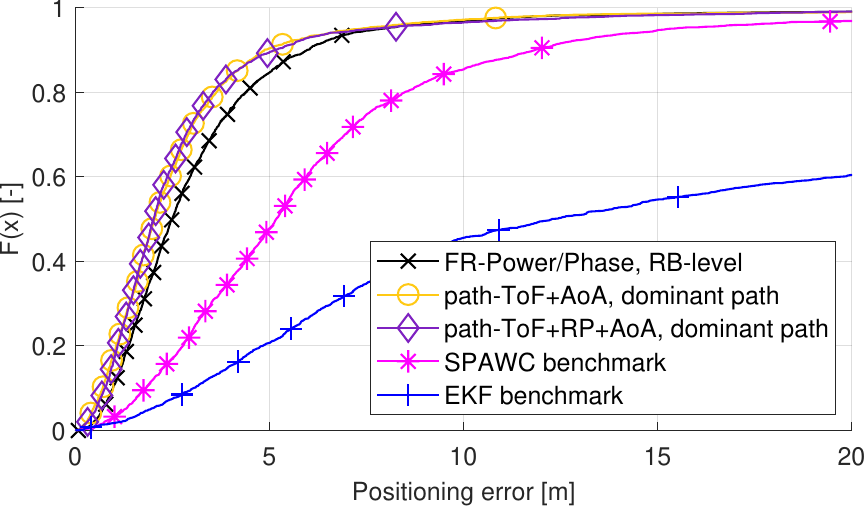}\label{fig:ecdf_pos_bench}}
  \hspace{2.0cm}
  \subfloat[]{\includegraphics[width=0.42\textwidth]{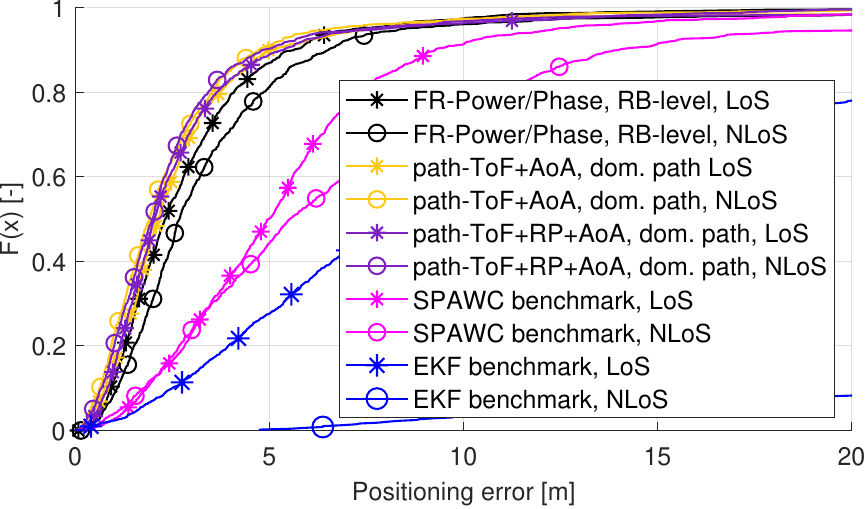}\label{fig:ecdf_pos_bench_los}}
  \vspace{-2mm}
  \caption{\acp{ecdf} of the positioning errors, in (a), and the split between the \ac{los} and \ac{nlos} samples, in (b), with realistic feature uncertainties. The proposed sequence processing \ac{nn} model with three different channel features is compared against two benchmark solutions.}
\end{figure*}

The positioning performance of the different models is visualized and compared in Fig.~\ref{fig:ecdf_pos_bench}, showing significantly improved positioning performance through the proposed solution compared to the two benchmarks. \red{While being evaluated with the same data, the proposed time- and frequency-domain \ac{csi} features combined with the proposed sequence processing engine achieve around $2$\,m median positioning error, compared to $5$\,m of the SPAWC benchmark and $12$\,m of the \ac{ekf} benchmark. The achieved performance improvement is stemming from the combination of the novel feature engineering and the carefully crafted sequence based NN processing system. Furthermore, besides estimating the UE location, incorporating the estimation of UE heading and velocity leads to reduced estimation variance through stabilization between individual quantities. Compared to the EKF model, which is built on the assumption of \ac{los} geometry and unbiased measurements, the proposed approach can work in both \ac{los} and \ac{nlos} conditions and deal with biased measurements -- such as the ones encountered with the discretized path-AoA feature.  }

Fig.~\ref{fig:ecdf_pos_bench_los} provides the corresponding results when distinguishing between the \ac{los} and \ac{nlos} regions. The \acp{ecdf} show that with the uncertainties within the received signals, the gap between the \ac{los} and \ac{nlos} performance with the proposed solution is strongly suppressed, as for the \ac{nn} model, the \ac{nlos} scenario itself represents an uncertainty on its own, which only complements the ones in the input data. While time-domain \ac{csi} features offer \ac{los}-agnostic results, there is still a small performance gap when utilizing the frequency-domain data. The SPAWC benchmark provides consistent results in the lower parts of the distribution, but the distribution for the \ac{nlos} samples contains significantly higher errors and outliers. The positioning performance of the \ac{ekf} in \ac{los} suffers from the discretization of the \ac{p-aoa}, leading to an increasing positioning error with the increasing distance from the \ac{los} \ac{gnb}(s). In \ac{nlos}, the \ac{ekf} follows the direction of the last \ac{los} sample, leading to unreliable estimation. {As noted already in the Introduction, there are Bayesian filtering or other model-based methods deliberately crafted for \ac{nlos} scenarios, such as \cite{ge2022computationally, talvitie2017novel}. However, such possess very high computational complexity and are typically limited to single-bounce scenarios, while in the evaluation environment considered in this article severe multi-bounce phenomena can occur.} 

\begin{figure*}[t!]
  \centering
  \subfloat[]{\includegraphics[width=0.42\textwidth]{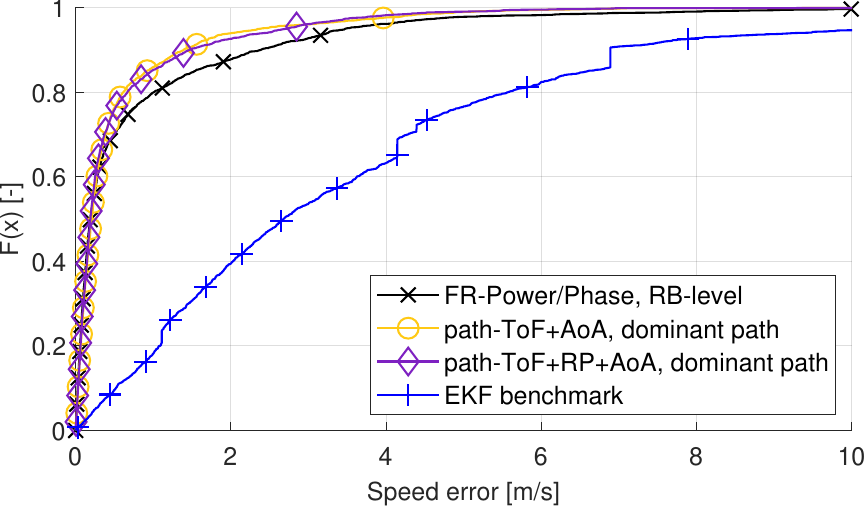}\label{fig:ecdf_speed_bench}}
  \hspace{2.0cm}
  \subfloat[]{\includegraphics[width=0.42\textwidth]{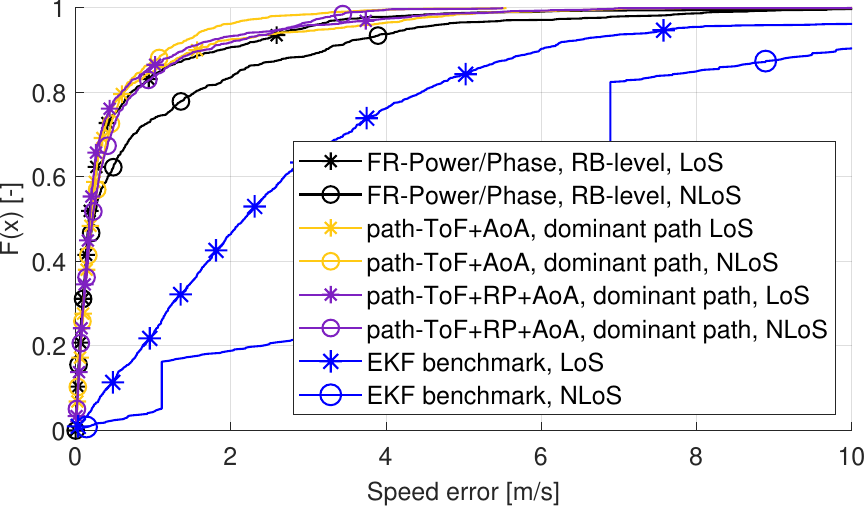}\label{fig:ecdf_speed_bench_los}}
  \vspace{-2mm}
  \caption{\acp{ecdf} of the speed estimation errors, in (a), and the split between the \ac{los} and \ac{nlos} samples, in (b), with realistic feature uncertainties. The proposed sequence processing \ac{nn} model with three different channel features is compared against the EKF benchmark solution.}
\end{figure*}

Fig.~\ref{fig:ecdf_speed_bench} and  Fig.~\ref{fig:ecdf_speed_bench_los} visualize the speed estimation and tracking performance in the same scenario as the positioning evaluation above, yet excluding the SPAWC benchmark which does not facilitate speed or heading estimation. The results show that the speed estimation error of the proposed model and features is in the majority of situations lower than $1$\,m/s, regardless of the input feature or \ac{los} availability. The frequency-domain feature \ac{fr-pp} has a slightly higher estimation error in \ac{nlos}. Comparably, \ac{ekf} is subject to significantly higher estimation error even in \ac{los}, while in \ac{nlos} loses the tracking capabilities completely.

\begin{figure*}[t!]
  \centering
  \vspace{-3mm}
  \subfloat[]{\includegraphics[width=0.42\textwidth]{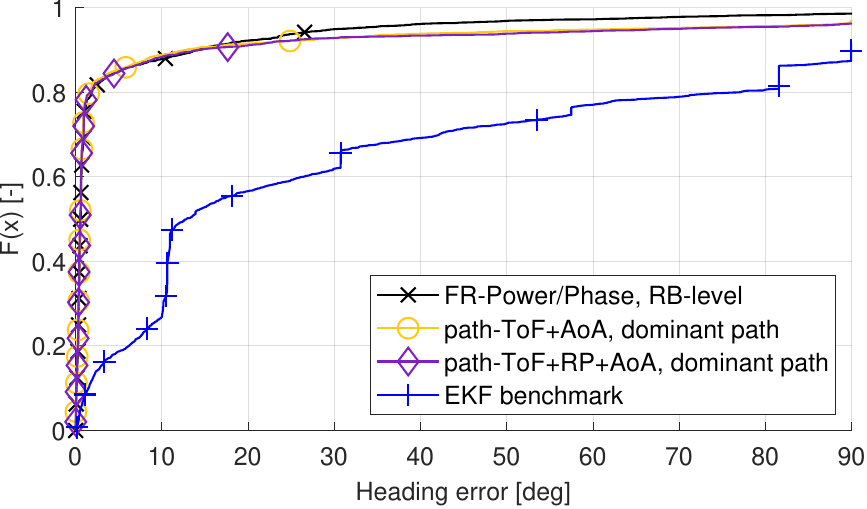}\label{fig:ecdf_head_bench}}
  \hspace{2.0cm}
  \subfloat[]{\includegraphics[width=0.42\textwidth]{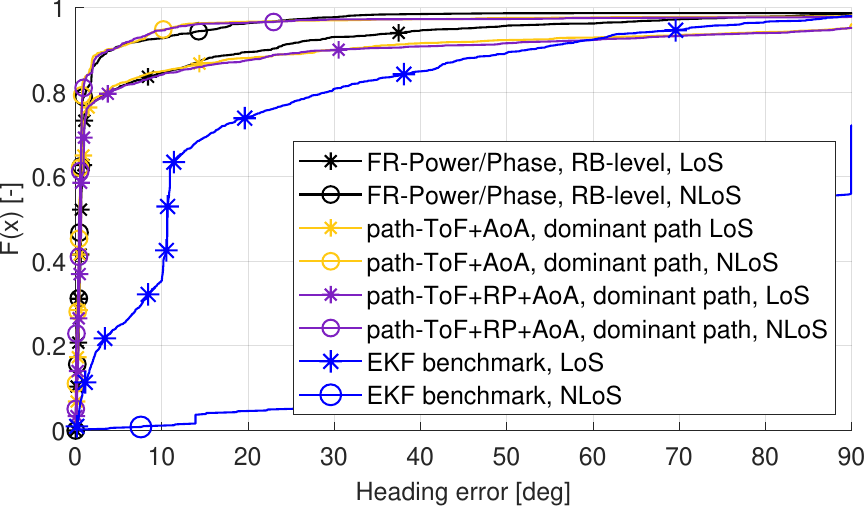}\label{fig:ecdf_head_bench_los}}
  \hspace{0.0cm}
  \vspace{-2mm}
  \caption{\acp{ecdf} of the heading estimation errors, in (a), and the split between the \ac{los} and \ac{nlos} samples, in (b), with realistic feature uncertainties.  The proposed sequence processing \ac{nn} model with three different channel features is compared against the EKF benchmark solution.}
  \vspace{-2.5mm}
\end{figure*}

Furthermore, as shown in Fig.~\ref{fig:ecdf_head_bench} and Fig.~\ref{fig:ecdf_head_bench_los}, the \ac{ue} heading is in the majority of samples estimated correctly -- with less than $1^\circ$ error in more than $75$\% of the samples. Interestingly, the \ac{nlos} heading estimates are even more accurate than the ones with an unobstructed radio link to the \ac{gnb}. This phenomenon occurs when the \ac{ue} starts turning and the heading rapidly changes, while the data, such as discretized \ac{p-aoa}, remain unaffected by the change. The model is not capable of instantaneously reacting due to the uncertainties included within the data. 

{
{For a comprehensive performance assessment}, we next shortly address the potential impact of the uncertainties in the training data positioning labels. We model the uncertainty through \ac{awgn} and set the corresponding \ac{std} in the x-y coordinates to $5$\,m. The obtained results are illustrated in Fig.~\ref{fig:boxplot_unc}. As can be observed, additional uncertainties in the labels force the model to generalize along the path. Consequently, the model is capable of tracking more effectively in \ac{los}, while in \ac{nlos}, the positioning results include an increased number of outliers.
These results show, that the proposed combination of features and the sequential NN model can diminish the effects of significant uncertainties in both features and labels.

\begin{figure}[t]
    \centering
    \includegraphics[width=0.99\columnwidth]{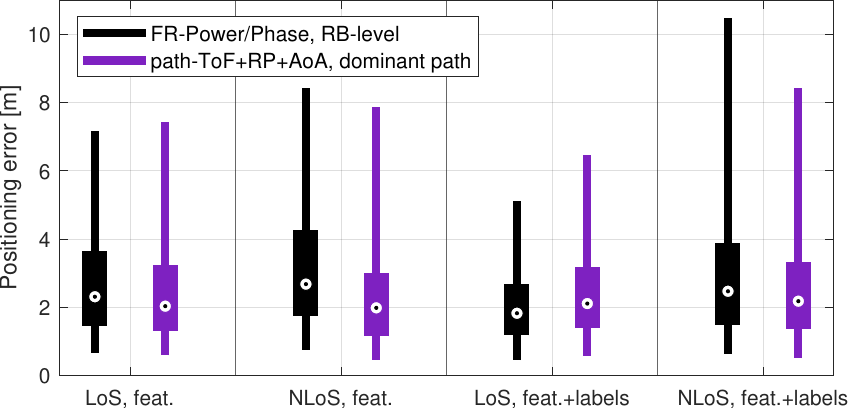}
    \captionsetup{belowskip=0pt}
    \vspace{-4mm}
    \caption{Distributions of the positioning errors with sequence-processing \ac{nn} model with feature uncertainties along the \ac{los} and \ac{nlos} regions, when considering perfect position labels (feat.) and the labels with the uncertainty modeled through AWGN with \ac{std} of $5$\,m along x-y coordinates (feat.\,+\,labels).}
    \vspace{-0mm}
    \label{fig:boxplot_unc}
\end{figure}
}

\vspace{-3mm}
\red{
\subsection{Deployment Specialization through Transfer Learning}

\textls[-1]{Being able to adapt the \ac{nn} model to environmental changes is a critical task in real-world applications, as any given model is essentially restricted to the environment it was trained at. This is especially so if no direct information about the environment, such as \ac{gnb} locations, beam directions, or buildings, is included in the feature vector. To address this practical challenge, we explore next the capabilities of \ac{tl}, which relies on re-training the model from one scenario to another. 
For presentation simplicity, we consider the snapshot model with \ac{fr-pp} features and assume no feature or label uncertainty. The models utilize the same training parameters as described in Section~\ref{sec:4nnmodels}, with the exception of applying the early stopping mechanism also to the first training loop. In addition, for the considered new \ac{gnb} deployments, we limit the dataset size to \emph{only $10\%$ of the original} 
to demonstrate the feasibility of the \ac{tl} approach also with small data sizes.}

\subsubsection{Scenario 1} \label{sec:5tl1}
We first consider \emph{relocating a single \ac{gnb}} within the deployment, thus altering the signal propagation geometry, as visualized in Fig.~\ref{fig:dep2}. For the evaluation, we consider three models, all sharing the same architecture: an original model, a \ac{tl} model, and a new model trained from scratch. The original model was trained on a full dataset from the prior deployment, whereas the \ac{tl} model and the new model were trained on the data from the altered scenario. Moreover, the newly trained model was initialized with random weights, while the \ac{tl} model was initialized with the weights of the original model before (re-)training. In addition, only the first layer after the input was set to ``trainable'', while the rest of the model remain frozen. 

The positioning error distributions of the considered models are visualized in Fig.~\ref{fig:TL_ecdf_s2}. Although the original model is not trained with the data from the altered deployment scenario, its positioning error is below $10$\,m in approximately~$40\%$ of the samples. However, most of the accurately localized samples of the original model are found in the southeast part of the deployment, where the relocated \ac{gnb} is not detected, and thus the environment seems essentially unchanged. The \ac{tl} model clearly outperforms the newly trained model, especially when considering errors above the $40^{th}$ percentile. The model convergence during the training is visualized in Fig.~\ref{fig:TL_training_s2}, where both training and validation losses across epochs are shown for the \ac{tl} and the newly trained models. While the \ac{tl} model is able to converge within less than $80$ epochs, the normally trained model requires more than $200$ epochs (not explicitly visible in the figure) to obtain the final weights. \blue{This highlights the training efficiency of the \ac{tl} model.}

\begin{figure*}[t!]
  \centering
  \vspace{-3mm}
  \hspace{0.2cm}
  \subfloat[]{\includegraphics[height=4.6cm]{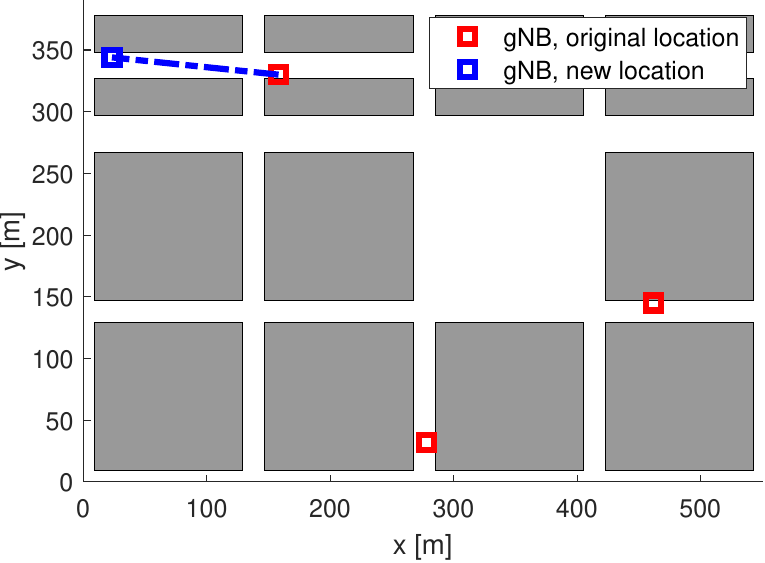}
  \label{fig:dep2}}
  \hfill
  \subfloat[]{\includegraphics[width=4.6cm]{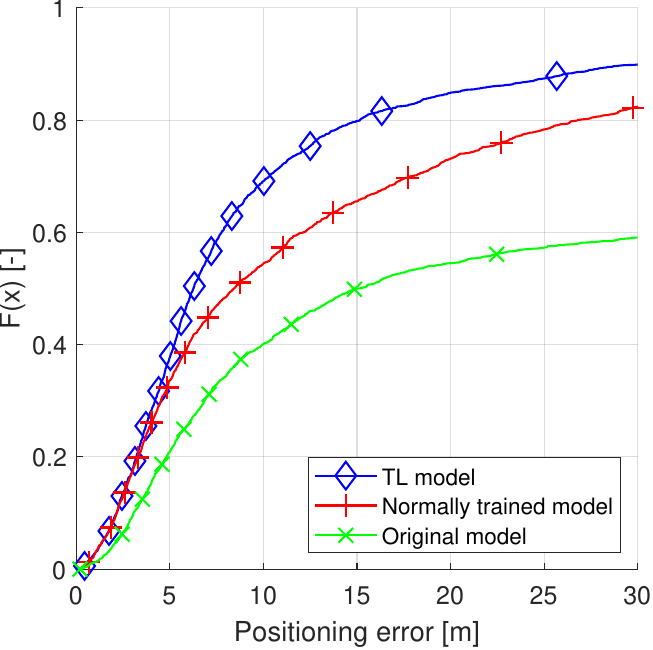}        
  \label{fig:TL_ecdf_s2}
  }
  \hfill
  \subfloat[]{\includegraphics[width=4.6cm]{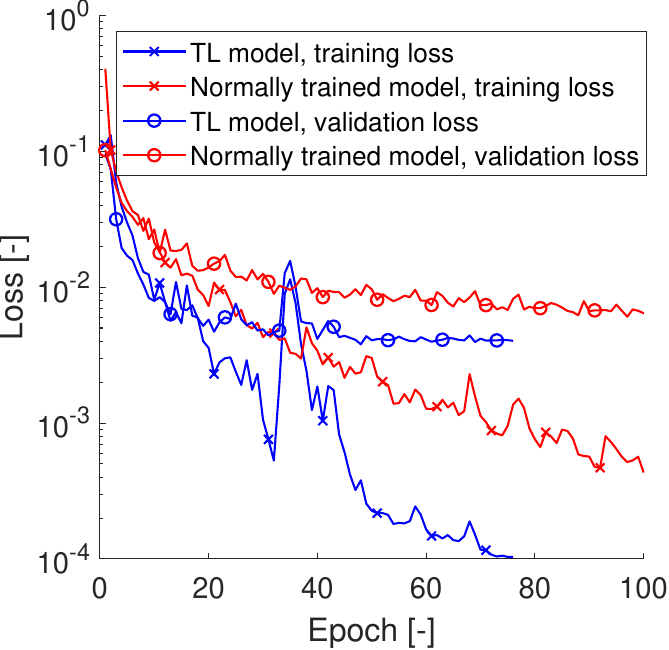}
  \label{fig:TL_training_s2}}
  \hfill
  \vspace{-1mm}
  \caption{\red{(a) Scenario 1 illustration where a single \ac{gnb} is relocated. (b) Positioning performance comparison between the model trained on the original data, adapted model via \ac{tl}, and a new model trained from scratch. (c) The training histories of the TL model and the newly trained model. 
  }
  }
  \vspace{-0mm}
\end{figure*}

\subsubsection{Scenario 2} \label{sec:5tl2}
We next consider the more challenging case of \emph{relocating all three \acp{gnb}}, as shown in Fig.~\ref{fig:dep3}. In this case, the changes in the \ac{gnb} coordinates are up to hundreds of meters, resulting in significantly altered radio propagation characteristics. In Fig.~\ref{fig:TL_ecdf_s3}, the positioning errors are shown following similar model cases as in Scenario~1. Due to the considerably changed \ac{gnb} locations, the performance of the original model clearly collapses.
Furthermore, the \ac{tl} model provides clearly better performance compared to the new model trained from scratch. In Fig.~\ref{fig:TL_training_s3}, it can be seen that besides providing the best positioning performance, the \ac{tl} model significantly accelerates the training rate, \blue{enabling fast model convergence in only some $50$ epochs}. These results show that \ac{tl} is an efficient approach to adapt the \ac{nn} model to a new, previously unseen scenario with greatly reduced effort and relaxed requirements on the availability of data. 

\begin{figure*}[t!]
  \centering
  \vspace{-3mm}
  \hspace{0.2cm}
  \subfloat[]{\includegraphics[height=4.6cm]{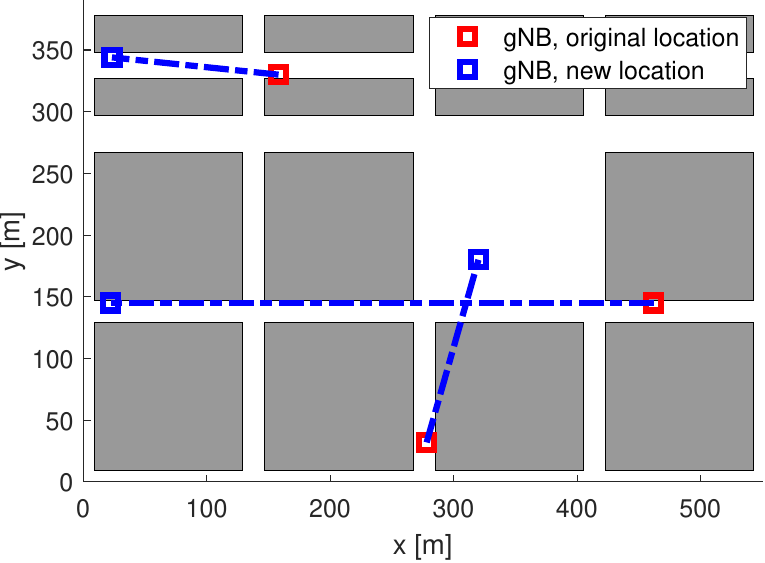}\label{fig:dep3}} 
  \hfill
  \subfloat[]{\includegraphics[width=4.6cm]{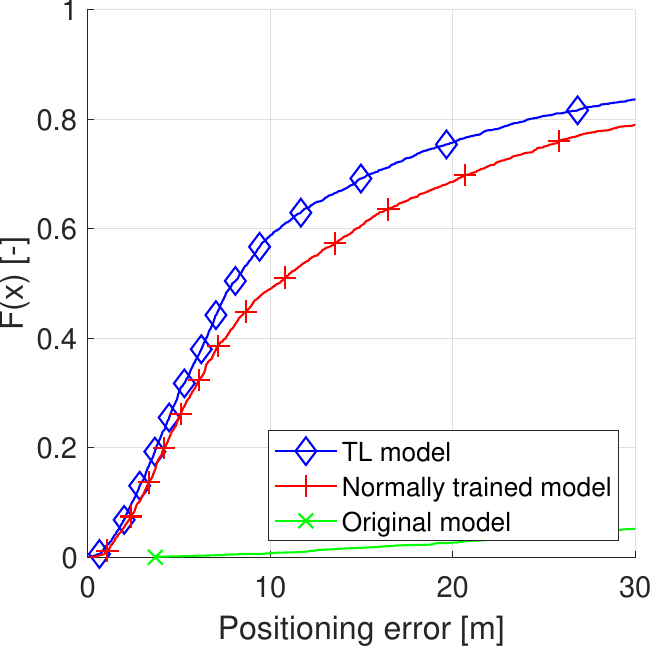}        
  \label{fig:TL_ecdf_s3}
  }
  \hfill
  \subfloat[]{\includegraphics[width=4.6cm]{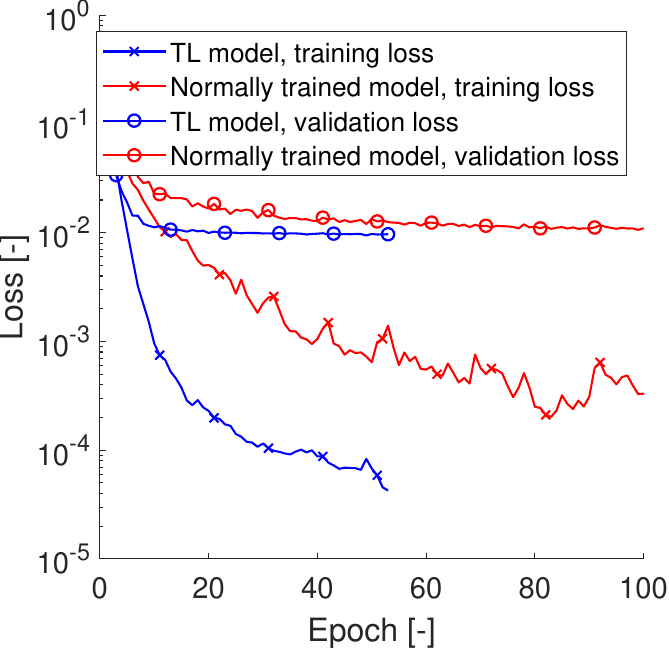}
  \label{fig:TL_training_s3}}
  \hfill
  \vspace{-1mm}
  \caption{\red{(a) Scenario 2 illustration where all \acp{gnb} are relocated. (b) Positioning performance comparison between the model trained on the original data, adapted model via \ac{tl}, and a new model trained from scratch. (c) The training histories of the TL model and the newly trained model.}}
  \vspace{-0mm}
\end{figure*}
}

\blue{
\subsection{Notes on Complexity}
\textls[-1]{Model complexity is an important aspect of any \ac{nn} solution, and notable efforts are commonly invested to reach beneficial performance-complexity trade-offs. 
For comparability and presentation simplicity, we focus here primarily on the parameter counts and input sizes as the main complexity-related metrics. To this end, Table~\ref{tab:complexity} compares the proposed solutions with the ones from the referred literature noting also the model structures.}

\begin{table}[t]
\caption{\textsc{\blue{Complexity assessment of related works}}}
\label{tab:complexity}
    \centering
    \begin{tabular}{c|c|c|c}
    \toprule
    \blue{\textbf{Reference}} & 
    \blue{\textbf{ML model}} & 
    \blue{\textbf{Input size}} & 
    \blue{\textbf{Num. parameters}} \\ 
    \midrule

\blue{\cite{8823059}}	&	\blue{CNN}	&	\blue{$18~432$}	& \blue{$13M$} \\
\blue{\cite{klus2021neural}}	&	\blue{DNN}	&	\blue{$224$}	& \blue{$2~331~650$} \\
\blue{\cite{gonultacs2021csi}}	&	\blue{DNN}	&	\blue{N/A}	& \blue{$2.5M$} \\
\blue{\cite{chen2017confi}}	&	\blue{CNN}	&	\blue{$2~700$}	& \blue{$9M$} \\
\blue{\cite{ferrand2020dnn}}	&	\blue{DNN}	&	\blue{$18~432$}	& \blue{$9~105~346$} \\
\blue{\cite{gao2022towards}}	&	\blue{CNN}	&	\blue{$15~440$}	& \blue{$2.6M$} \\
\toprule
    \blue{\textbf{Our models}} & 
    \blue{\textbf{Feature}} & 
    \blue{\textbf{Input size}} & 
    \blue{\textbf{Num. parameters}} \\ 
    \midrule
\blue{\,Snapshot\,}	&	\blue{\,path-ToF+AoA\,}	&	\blue{$12$} & \blue{$1~058~306$} \\
\blue{\,Snapshot\,}	&	\blue{\,path-ToF+RP+AoA\,}	&	\blue{$15$} & \blue{$1~067~010$} \\
\blue{\,Snapshot\,}	&	\blue{\,FR-Power/Phase\,}	&	\blue{$960$} & \blue{$1~543~682$} \\
\blue{\,Sequence\,}	&	\blue{\,path-ToF+AoA\,}	&	\blue{$50\times12$*} & \blue{$1~667~192$} \\
\blue{\,Sequence\,}	&	\blue{\,path-ToF+RP+AoA\,}	&	\blue{$50\times15$*} & \blue{$1~670~264$} \\
\blue{\,Sequence\,}	&	\blue{\,FR-Power/Phase\,}	&	\blue{$50\times960$*} & \blue{$2~637~944$} \\
\bottomrule

\end{tabular}
\\
\vspace{1mm}
{ \blue{*\,Processing $50$ snapshot samples in a sequence}}
\vspace{-2mm}
\end{table}

As can be observed, both the convolutional and fully connected models utilized across the referred works consider a few million parameter models with up to $18~432$ input sizes, corresponding to manageable computational complexity when training on high-performance machines. In comparison, the path-ToF+AoA model which achieved meter-level positioning accuracies requires only $12$ inputs and a model with around one million parameters, while increasing the input size to $960$ with \ac{fr-pp} feature results in $1~543~682$ trainable parameters. Our sequence-processing models, despite their extra layers, retain relatively low parameter counts resulting in feasible computational requirements. Overall, we can conclude that while outperforming the prior-art methods in positioning accuracy, the proposed models are also computationally feasible -- and even lighter in complexity compared to many reference models.
}

\vspace{-2mm}
\subsection{\blue{Discussion on Synthetic vs. Real-World Training Data}}

While we evaluate the performance of the methods and models on ray-tracing data in this article, training the models with artificially created synthetic data can be a feasible option also for true deployments and applications of large-scale \ac{nn}-operated systems. In general, the acquisition of synthetic data is cheaper and faster than performing exhaustive site surveys and can approximate reality with increasing accuracy and fidelity~\cite{assayag2021indoor}. \red{Considering the availability of synthetic data, there are many possibilities to organize the model training in practice. For example, initial training can be performed already at the factory based on synthetic data from a specific intended operation environment, and then the model can be fine-tuned on a limited set of real-world measurements. \blue{Such real-world labelled data can be obtained through different crowdsourcing or crowdsensing arrangements, as discussed further below.} On the other hand, training at the factory can be more generic and cover numerous environments, while specialization to a specific environment can be managed with a \ac{tl} approach using synthetic data and/or real-world measurements. Similarly, in case of any changes in the environment, it is possible to update the model via \ac{tl} by utilizing re-generated site-specific synthetic data and/or newly obtained real-world measurements.}

\textls[-2]{\blue{Employing \emph{crowdsourcing} campaigns, where the surveying software is offered to the public to perform the measurements is one way of arranging real-world labelled data in practice. One recognized challenge is that such an approach is often biased by the human factor, such as inaccurate manual labeling. As another alternative, while crowdsourcing requires the user to perform an action to obtain the data, \emph{crowdsensing} is fully automated and unsupervised by the user. Consequently, it can yield a massive volume of data, though at the cost of potentially missing labels or other quality-related challenges. Numerous solutions exist to cure and filter such data, although some challenges still remain~\cite{capponi2019survey}.}}

In general, the topic of synthetic data is currently explored by IEEE~\cite{Synthetic_2023} and can become a critical link enabling real-world systems driven by their digital twins~\cite{castellani2020real}. This is a key component in positioning-driven studies, where obtaining realistic performance evaluations requires consideration of whole network deployments with detailed physical-layer processing and measurement capabilities. 
End-to-end simulated results play thus a crucial role in the positioning system analysis before conclusive validation through experimental field tests.

\vspace{-0mm}
\section{Conclusions}
\vspace{-0mm}
\label{sec:6}

\textls[-2]{This article addressed cellular network-based user localization and tracking in challenging \ac{nlos} environments, with specific emphasis on \ac{5g} \ac{mmw} deployments and urban vehicular systems. We first described the \ac{ul}/\ac{dl} measurements, available for positioning purposes, together with their acquisition in \ac{5g} \ac{mmw} networks. We then derived and proposed efficient frequency-domain \ac{csi} features, most notably utilizing the relative phases and powers of the received signal across the neighboring resource blocks. As time-domain \ac{csi} data, we exploit the multipath components and proposed different aggregate features combining time-of-flight, angle-of-arrival, and received path-wise powers. Deep learning \ac{ml} architectures were then described, covering not only dense snapshot models but also sequence-processing \ac{nn} models harnessing the temporal correlations of the features in vehicular systems.}

\textls[-1]{Realistic numerical evaluations in large-scale \ac{los}-obstructed urban environment with moving vehicles were provided, building on full ray-tracing-based propagation modeling on METIS Madrid map at 28\,GHz. The baseline results without feature uncertainties show that the frequency-domain \ac{csi} in the form of RB-level relative phases and powers allows for very good and robust positioning performance, in both \ac{los} and \ac{nlos}, while even further enhanced performance can be obtained through the time-domain features when combining multipath times-of-flight and angles-of-arrival. The results also show that dominant multipath feature combinations are sufficient, or even favorable, for robust positioning. Additionally, when considering practical levels of feature measurement uncertainties, together with the sequence-processing \ac{nn} models, robust positioning in both \ac{los} and \ac{nlos} was still shown to be feasible. 
\red{Finally, the important practical aspect of dealing with \ac{gnb} deployment differences between the training and inference phases was addressed. It was shown that such environment related uncertainties can be addressed and alleviated through transfer learning.}}

{Overall, the provided numerical results clearly demonstrate that the proposed methods harnessing the novel feature engineering and sequence processing neural network models outperform the state-of-the-art, being able to facilitate 1-2\,m median positioning accuracy even in deep-NLoS regions \blue{with feasible parameter counts}.} 
Our future work will focus on exploring the opportunities with obtaining and processing positioning measurements from co-existing C-band and \ac{mmw} cellular networks, as well as with further improving the positioning accuracy and reliability through sensor fusion.

\balance
\bibliographystyle{IEEEtran}
{\footnotesize
\bibliography{bibliography.bib}
}

\end{document}